\documentclass[aps,prd,twocolumn,showpacs,superscriptaddress]{revtex4-1}
\pdfoutput=1
\usepackage {amsmath}
\usepackage {amssymb}
\usepackage {amsfonts}
\usepackage {amsthm}
\usepackage {mathrsfs}
\usepackage {natbib}
\usepackage {latexsym}
\usepackage {graphicx}
\usepackage {dsfont}
\usepackage {txfonts}
\usepackage {rotating}
\usepackage {wasysym}
\usepackage {multirow}
\usepackage {hhline}
\usepackage {graphicx}
\usepackage {hyperref}
\usepackage {color}
\usepackage {bm}
\usepackage {appendix}
\usepackage {acronym}
\usepackage {dcolumn}   
\usepackage {subfigure}
\usepackage {url}
\usepackage {xspace}

\usepackage[latin1]{inputenc}
\usepackage{tikz}
\usetikzlibrary{shapes,arrows,shapes.multipart}

\usepackage{verbatim}

\newcommand{\dcc}{LIGO-P1400094}

\newcommand{\dg}{^\circ} 
\newcommand{\Msol}{M_\odot}
\newcommand{\Mstar}{M}		

\newcommand{\fstat}{\mathcal{F}} 
\newcommand{\cstat}{\mathcal{C}}
\newcommand{\ftext}{$\mathcal{F}$\xspace}
\newcommand{\ctext}{$\mathcal{C}$\xspace}
\newcommand{\fstext}{{$\mathcal{F}$}-statistic\xspace}
\newcommand{\cstext}{{$\mathcal{C}$}-statistic\xspace}

\newcommand{\cstexts}{{$\mathcal{C}$}-statistics\xspace}
\newcommand{\cstar}{\cstat^*}

\newcommand{\subtext}[1]{\text{\tiny{#1}}}
\newcommand{\submath}[1]{\text{\tiny{\emph{#1}}}}

\newcommand{\Tobs}{T_\text{obs}}		
\newcommand{\Ts}{T_\text{s}}			
\newcommand{\Fx}{F_\subtext{X}} 		
\newcommand{\fspin}{\nu_\text{s}}		
\newcommand{\fo}{f_\submath{0}}			
\newcommand{\ho}{h_\submath{0}}			
\newcommand{\pUL}{p_\subtext{UL}}		
\newcommand{\hUL}{h_\subtext{UL}}		
\newcommand{\hULn}{h_\subtext{UL}^{95}}		
\newcommand{\epUL}{\epsilon_\subtext{UL}}	
\newcommand{\Porb}{P_\text{orb}}		
\newcommand{\asini}{a_\submath{0}}		
\newcommand{\Pa}{P_\text{a}}			
\newcommand{\PaN}{ P_{\text{a}|N}}		

\tikzstyle{decision} = [diamond, draw, fill=green!20, rounded corners=1pt, 
    text width=5em, text badly centered, node distance=3cm, inner sep=0pt]
\tikzstyle{process} = [rectangle, draw, fill=blue!20, 
    text width=8em, text centered, rounded corners=1pt, minimum height=4em]
\tikzstyle{line} = [draw, -latex']
\tikzstyle{terminal} = [draw, ellipse,fill=red!20, node distance=3cm, minimum
height=2em]
\tikzstyle{data} = [draw,trapezium,trapezium left angle=70,trapezium right
angle=-70, text width=2.5cm, text centered, rounded corners=1pt, node
distance=3cm, minimum height=1em]

\begin{document}



\title{A directed search for gravitational waves from Scorpius X-1 with initial LIGO} 



\author{%
J.~Aasi,$^{1}$
B.~P.~Abbott,$^{1}$
R.~Abbott,$^{1}$
T.~Abbott,$^{2}$
M.~R.~Abernathy,$^{1}$
F.~Acernese,$^{3,4}$
K.~Ackley,$^{5}$
C.~Adams,$^{6}$
T.~Adams,$^{7,8}$
T.~Adams,$^{8}$
% ...
}\noaffiliation

\affiliation {LIGO, California Institute of Technology, Pasadena, CA 91125, USA }
\affiliation {Louisiana State University, Baton Rouge, LA 70803, USA }
\affiliation {Universit\`a di Salerno, Fisciano, I-84084 Salerno, Italy }
\affiliation {INFN, Sezione di Napoli, Complesso Universitario di Monte Sant'Angelo, I-80126 Napoli, Italy }
\affiliation {University of Florida, Gainesville, FL 32611, USA }
\affiliation {LIGO Livingston Observatory, Livingston, LA 70754, USA }
\affiliation {Cardiff University, Cardiff, CF24 3AA, United Kingdom }
\affiliation {Laboratoire d'Annecy-le-Vieux de Physique des Particules (LAPP), Universit\'e de Savoie, CNRS/IN2P3, F-74941 Annecy-le-Vieux, France }
\affiliation {University of Sannio at Benevento, I-82100 Benevento, Italy and INFN, Sezione di Napoli, I-80100 Napoli, Italy }
\affiliation {Experimental Group, Albert-Einstein-Institut, Max-Planck-Institut f\"ur Gravi\-ta\-tions\-physik, D-30167 Hannover, Germany }
\affiliation {Nikhef, Science Park, 1098 XG Amsterdam, The Netherlands }
\affiliation {LIGO, Massachusetts Institute of Technology, Cambridge, MA 02139, USA }
\affiliation {Instituto Nacional de Pesquisas Espaciais, 12227-010 S\~{a}o Jos\'{e} dos Campos, SP, Brazil }
\affiliation {Inter-University Centre for Astronomy and Astrophysics, Pune 411007, India }
\affiliation {International Centre for Theoretical Sciences, Tata Institute of Fundamental Research, Bangalore 560012, India }
\affiliation {Syracuse University, Syracuse, NY 13244, USA }
\affiliation {Data Analysis Group, Albert-Einstein-Institut, Max-Planck-Institut f\"ur Gravitations\-physik, D-30167 Hannover, Germany }
\affiliation {University of Wisconsin--Milwaukee, Milwaukee, WI 53201, USA }
\affiliation {Universit\`a di Siena, I-53100 Siena, Italy }
\affiliation {INFN, Sezione di Pisa, I-56127 Pisa, Italy }
\affiliation {The University of Mississippi, University, MS 38677, USA }
\affiliation {California State University Fullerton, Fullerton, CA 92831, USA }
\affiliation {LAL, Universit\'e Paris-Sud, IN2P3/CNRS, F-91898 Orsay, France }
\affiliation {University of Southampton, Southampton, SO17 1BJ, United Kingdom }
\affiliation {Leibniz Universit\"at Hannover, D-30167 Hannover, Germany }
\affiliation {INFN, Sezione di Roma, I-00185 Roma, Italy }
\affiliation {University of Birmingham, Birmingham, B15 2TT, United Kingdom }
\affiliation {Albert-Einstein-Institut, Max-Planck-Institut f\"ur Gravitations\-physik, D-14476 Golm, Germany }
\affiliation {Montana State University, Bozeman, MT 59717, USA }
\affiliation {Universit\`a di Perugia, I-06123 Perugia, Italy }
\affiliation {INFN, Sezione di Perugia, I-06123 Perugia, Italy }
\affiliation {European Gravitational Observatory (EGO), I-56021 Cascina, Pisa, Italy }
\affiliation {SUPA, University of Glasgow, Glasgow, G12 8QQ, United Kingdom }
\affiliation {LIGO Hanford Observatory, Richland, WA 99352, USA }
\affiliation {APC, AstroParticule et Cosmologie, Universit\'e Paris Diderot, CNRS/IN2P3, CEA/Irfu, Observatoire de Paris, Sorbonne Paris Cit\'e, 10, rue Alice Domon et L\'eonie Duquet, F-75205 Paris Cedex 13, France }
\affiliation {Columbia University, New York, NY 10027, USA }
\affiliation {Stanford University, Stanford, CA 94305, USA }
\affiliation {Universit\`a di Pisa, I-56127 Pisa, Italy }
\affiliation {CAMK-PAN, 00-716 Warsaw, Poland }
\affiliation {Astronomical Observatory Warsaw University, 00-478 Warsaw, Poland }
\affiliation {The University of Texas at Brownsville, Brownsville, TX 78520, USA }
\affiliation {Universit\`a degli Studi di Genova, I-16146 Genova, Italy }
\affiliation {INFN, Sezione di Genova, I-16146 Genova, Italy }
\affiliation {RRCAT, Indore MP 452013, India }
\affiliation {Faculty of Physics, Lomonosov Moscow State University, Moscow 119991, Russia }
\affiliation {NASA/Goddard Space Flight Center, Greenbelt, MD 20771, USA }
\affiliation {University of Western Australia, Crawley, WA 6009, Australia }
\affiliation {Department of Astrophysics/IMAPP, Radboud University Nijmegen, P.O. Box 9010, 6500 GL Nijmegen, The Netherlands }
\affiliation {ARTEMIS, Universit\'e Nice-Sophia-Antipolis, CNRS and Observatoire de la C\^ote d'Azur, F-06304 Nice, France }
\affiliation {MTA E\"otv\"os University, `Lendulet' Astrophysics Research Group, Budapest 1117, Hungary }
\affiliation {Institut de Physique de Rennes, CNRS, Universit\'e de Rennes 1, F-35042 Rennes, France }
\affiliation {Washington State University, Pullman, WA 99164, USA }
\affiliation {Universit\`a degli Studi di Urbino 'Carlo Bo', I-61029 Urbino, Italy }
\affiliation {INFN, Sezione di Firenze, I-50019 Sesto Fiorentino, Firenze, Italy }
\affiliation {University of Oregon, Eugene, OR 97403, USA }
\affiliation {Laboratoire Kastler Brossel, ENS, CNRS, UPMC, Universit\'e Pierre et Marie Curie, F-75005 Paris, France }
\affiliation {VU University Amsterdam, 1081 HV Amsterdam, The Netherlands }
\affiliation {University of Maryland, College Park, MD 20742, USA }
\affiliation {University of Massachusetts Amherst, Amherst, MA 01003, USA }
\affiliation {Laboratoire des Mat\'eriaux Avanc\'es (LMA), IN2P3/CNRS, Universit\'e de Lyon, F-69622 Villeurbanne, Lyon, France }
\affiliation {Universitat de les Illes Balears---IEEC, E-07122 Palma de Mallorca, Spain }
\affiliation {Universit\`a di Napoli 'Federico II', Complesso Universitario di Monte Sant'Angelo, I-80126 Napoli, Italy }
\affiliation {Canadian Institute for Theoretical Astrophysics, University of Toronto, Toronto, Ontario, M5S 3H8, Canada }
\affiliation {Tsinghua University, Beijing 100084, China }
\affiliation {University of Michigan, Ann Arbor, MI 48109, USA }
\affiliation {INFN, Sezione di Roma Tor Vergata, I-00133 Roma, Italy }
\affiliation {National Tsing Hua University, Hsinchu Taiwan 300 }
\affiliation {Charles Sturt University, Wagga Wagga, NSW 2678, Australia }
\affiliation {Caltech-CaRT, Pasadena, CA 91125, USA }
\affiliation {Pusan National University, Busan 609-735, Korea }
\affiliation {Australian National University, Canberra, ACT 0200, Australia }
\affiliation {Carleton College, Northfield, MN 55057, USA }
\affiliation {Universit\`a di Roma Tor Vergata, I-00133 Roma, Italy }
\affiliation {INFN, Gran Sasso Science Institute, I-67100 L'Aquila, Italy }
\affiliation {Universit\`a di Roma 'La Sapienza', I-00185 Roma, Italy }
\affiliation {University of Brussels, Brussels 1050, Belgium }
\affiliation {Sonoma State University, Rohnert Park, CA 94928, USA }
\affiliation {Texas Tech University, Lubbock, TX 79409, USA }
\affiliation {University of Minnesota, Minneapolis, MN 55455, USA }
\affiliation {The University of Sheffield, Sheffield S10 2TN, United Kingdom }
\affiliation {Wigner RCP, RMKI, H-1121 Budapest, Konkoly Thege Mikl\'os \'ut 29-33, Hungary }
\affiliation {Montclair State University, Montclair, NJ 07043, USA }
\affiliation {Argentinian Gravitational Wave Group, Cordoba Cordoba 5000, Argentina }
\affiliation {Universit\`a di Trento, I-38123 Povo, Trento, Italy }
\affiliation {INFN, Trento Institute for Fundamental Physics and Applications, I-38123 Povo, Trento, Italy }
\affiliation {Northwestern University, Evanston, IL 60208, USA }
\affiliation {University of Cambridge, Cambridge, CB2 1TN, United Kingdom }
\affiliation {University of Szeged, D\'om t\'er 9, Szeged 6720, Hungary }
\affiliation {Rutherford Appleton Laboratory, HSIC, Chilton, Didcot, Oxon, OX11 0QX, United Kingdom }
\affiliation {Embry-Riddle Aeronautical University, Prescott, AZ 86301, USA }
\affiliation {The Pennsylvania State University, University Park, PA 16802, USA }
\affiliation {American University, Washington, DC 20016, USA }
\affiliation {University of Adelaide, Adelaide, SA 5005, Australia }
\affiliation {West Virginia University, Morgantown, WV 26506, USA }
\affiliation {Raman Research Institute, Bangalore, Karnataka 560080, India }
\affiliation {Korea Institute of Science and Technology Information, Daejeon 305-806, Korea }
\affiliation {University of Bia{\l }ystok, 15-424 Bia{\l }ystok, Poland }
\affiliation {SUPA, University of Strathclyde, Glasgow, G1 1XQ, United Kingdom }
\affiliation {IISER-TVM, CET Campus, Trivandrum Kerala 695016, India }
\affiliation {Institute of Applied Physics, Nizhny Novgorod, 603950, Russia }
\affiliation {Seoul National University, Seoul 151-742, Korea }
\affiliation {Hanyang University, Seoul 133-791, Korea }
\affiliation {NCBJ, 05-400 \'Swierk-Otwock, Poland }
\affiliation {IM-PAN, 00-956 Warsaw, Poland }
\affiliation {Institute for Plasma Research, Bhat, Gandhinagar 382428, India }
\affiliation {The University of Melbourne, Parkville, VIC 3010, Australia }
\affiliation {INFN, Sezione di Padova, I-35131 Padova, Italy }
\affiliation {Monash University, Victoria 3800, Australia }
\affiliation {ESPCI, CNRS, F-75005 Paris, France }
\affiliation {Universit\`a di Camerino, Dipartimento di Fisica, I-62032 Camerino, Italy }
\affiliation {Southern University and A\&M College, Baton Rouge, LA 70813, USA }
\affiliation {College of William and Mary, Williamsburg, VA 23187, USA }
\affiliation {Abilene Christian University, Abilene, TX 79699, USA }
\affiliation {Instituto de F\'\i sica Te\'orica, University Estadual Paulista/ICTP South American Institute for Fundamental Research, S\~ao Paulo SP 01140-070, Brazil }
\affiliation {IISER-Kolkata, Mohanpur, West Bengal 741252, India }
\affiliation {Whitman College, 280 Boyer Ave, Walla Walla, WA 9936, USA }
\affiliation {National Institute for Mathematical Sciences, Daejeon 305-390, Korea }
\affiliation {Rochester Institute of Technology, Rochester, NY 14623, USA }
\affiliation {Hobart and William Smith Colleges, Geneva, NY 14456, USA }
\affiliation {Tata Institute for Fundamental Research, Mumbai 400005, India }
\affiliation {SUPA, University of the West of Scotland, Paisley, PA1 2BE, United Kingdom }
\affiliation {Institute of Astronomy, 65-265 Zielona G\'ora, Poland }
\affiliation {Universit\"at Hamburg, D-22761 Hamburg, Germany }
\affiliation {Indian Institute of Technology, Gandhinagar Ahmedabad Gujarat 382424, India }
\affiliation {Andrews University, Berrien Springs, MI 49104, USA }
\affiliation {Trinity University, San Antonio, TX 78212, USA }
\affiliation {University of Washington, Seattle, WA 98195, USA }





\date{\today \\ \mbox{\dcc}}

\begin{abstract}
We present results of a search for continuously-emitted gravitational radiation, directed at the brightest low-mass X-ray binary, Scorpius X-1. Our semi-coherent analysis covers 10 days of LIGO S5 data ranging from 50--550 Hz, and performs an incoherent sum of coherent \ftext-statistic power distributed amongst frequency-modulated orbital sidebands. All candidates not removed at the veto stage were found to be consistent with noise at a $1\%$ false alarm rate. We present Bayesian 95\% confidence upper limits on gravitational-wave strain amplitude using two different prior distributions: a standard one, with no a priori assumptions about the orientation of Scorpius X-1; and an angle-restricted one, using a prior derived from electromagnetic observations. Median strain upper limits of $1.3\times 10^{-24}$ and $8\times 10^{-25}$ are reported at 150 Hz for the standard and angle-restricted searches respectively. This proof of principle analysis was limited to a short observation time by unknown effects of accretion on the intrinsic spin frequency of the neutron star, but improves upon previous upper limits by factors of ${\sim} 1.4$ for the standard, and 2.3 for the angle-restricted search at the sensitive region of the detector.
\end{abstract}



\maketitle

\acrodef{pdf}[pdf]{probability density function}
\acrodef{CDF}[CDF]{cumulative distribution function}
\acrodef{GW}[GW]{gravitational wave}
\acrodef{EM}[EM]{electromagnetic}
\acrodef{SNR}[SNR]{signal-to-noise-ratio}
\acrodef{NS}[NS]{neutron star}
\acrodef{LMXB}[LMXB]{low-mass X-ray binary}
\acrodef{QPO}{quasi-periodic oscillation}
\acrodef{ScoX1}[Sco X-1]{Scorpius X-1}
\acrodef{LIGO}{Laser Interferometer Gravitational Wave Observatory}
\acrodef{LSC}{LIGO Scientific Collaboration}
\acrodef{SFT}{short Fourier transform}
\acrodef{MC}{Monte Carlo}
\acrodef{RMS}{root-mean-square}
\acrodef{eos}[EOS]{equation of state}

\section{Introduction}\label{sec:intro}

Recycled neutron stars are a likely source of persistent,
quasi-monochromatic gravitational waves detectable by ground-based
interferometric detectors. Emission mechanisms include
thermocompositional and magnetic
mountains~\cite{Bonazzola_Gourgoulhon_1996,UCB_2000,Melatos_Payne_2005,Vigelius_Melatos_2009,PriymakEA_2011,HaskellEA_2006},
unstable oscillation modes~\cite{OwenEA_1998} and free
precession~\cite{Jones_2002}. If the angular momentum lost to
gravitational radiation is balanced by the spin-up torque from
accretion, the gravitational wave strain $\ho$ can be estimated
independently of the microphysical origin of the quadrupole and is
proportional to the observable X-ray
flux $F_x$ and spin frequency $\fspin$ ~\cite{Wagoner_1984,Bildsten_1998} via $\ho \propto \left({\Fx}/{\fspin}\right)^{1/2}$. Given the assumption of torque balance, the strongest
gravitational wave sources are those that are most proximate with the
highest accretion rate and hence X-ray flux, such as \ac{LMXB}
systems. In this sense the most luminous gravitational wave \ac{LMXB}
source is \ac{ScoX1}. 

The plausibility of the torque-balance scenario
is strengthened by observations of the spin frequencies
$\fspin$ of pulsating or bursting \acp{LMXB},
which show them clustered in a relatively narrow band from $270
\leqslant \fspin \leqslant 620$ Hz~\cite{ChakrabartyEA_Nature_2003},
even though their ages and accretions rates imply that they should
have accreted enough matter to reach the centrifugal break-up limit
$\nu_{\text{max}}{\sim} 1400$ Hz~\cite{CookEA_1994} of the neutron
star. The gravitational wave spin-down torque scales as $\fspin^5$, mapping a wide range of accretion rates into a narrow range of
equilibrium spins, so far conforming with observations. Alternative explanations for the clustering of \ac{LMXB} spin periods involving disc accretion physics have been proposed \cite{HaskellEA_2014}. Although this explanation suggests that gravitational radiation is not required to brake the spin-up of the neutron star, it does not rule out gravitational emission from these systems. The gravitational-wave torque-balance argument is used here as an approximate bound.

The initial instruments installed in the \ac{LIGO} consisted of three Michelson interferometers, one with 4-km orthogonal arms at Livingston, LA, and two collocated at Hanford, CA, with 4 km and 2 km arms. Initial \ac{LIGO} achieved its design
sensitivity during its fifth science run (between November 2005 and
October 2007 )~\cite{LIGO_2009,LIGO_S5cal_Abadie2010} and is currently
being upgraded to the next-generation Advanced \ac{LIGO}
configuration, which is expected to improve its sensitivity
ten-fold in strain~\cite{aLIGO_Harry_2010}.

Three types of searches have previously been conducted with \ac{LIGO} data for \ac{ScoX1}. The first, a coherent analysis using data from \ac{LIGO}'s
second science run (S2), was computationally limited to six-hour data
segments. It placed a wave-strain upper limit at $95 \%$ confidence of
$h_0^{95} \approx 2\times 10^{-22}$ for two 20 Hz bands between 464 --
484 Hz and 604 -- 626 Hz~\cite{LIGO_S2_ScoX1_2007}. The second, employing a radiometer technique~\cite{Ballmer_2006}, was conducted using all 20 days of \ac{LIGO} S4 data~\cite{LIGO_S4_ScoX1_2007}. It improved the upper limits on the
previous (S2) search by an order of magnitude in the relevant
frequency bands but did not yield a detection. The same method was
later applied to S5 data and reported roughly a five-fold sensitivity
improvement over the S4 results~\cite{LIGO_S5_ScoX1stoch_2011}. The S5
analysis returned a median $90\%$ confidence root-mean-square strain upper
limit of $h^{90}_{\text{rms}}=7\times 10^{-25}$ at 150 Hz, the most
sensitive detector frequency (this converts to $h_0^{95} \approx 2\times10^{-24}$ \cite{strain,rad,Messenger2010}). Thirdly, an all-sky search for continuous gravitational waves from sources in binary systems, which looks for patterns caused by binary orbital motion doubly Fourier-transformed data (TwoSpect), was adapted to search the \ac{ScoX1} sky position, and returned results in the low frequency band from $20 - 57.25$ Hz ~\cite{LIGO_TwoSpect_2014}.

Here we implement a new search for gravitational waves from sources in
known binary systems, with unknown spin frequency, initially directed
at \ac{ScoX1} on \ac{LIGO} S5 data to demonstrate feasibility. Values of the coherent, matched-filtered \ftext-statistic~\cite{JKS1998} are incoherently summed at the locations of frequency-modulated sidebands. This multi-stage, semi-coherent, analysis yields a new detection statistic, denoted the \ctext-statistic~\cite{MW2007, SammutEA_2014}. A similar technique was first employed in electromagnetic searches for radio pulsars~\cite{RansomEA_2003}. We utilise this technique to efficiently deal with the large parameter space introduced by the orbital motion of a source in a binary system.

A brief description of the search is given in Sec. \ref{sec:sideband}, while the astrophysical target source and its associated parameter space are discussed in Sec. \ref{sec:params}. Section \ref{sec:implement} outlines the search method, reviews
the pipeline, discusses the selection and preprocessing of \ac{LIGO}
S5 data, and explains the post-processing procedure. Results of the search, including upper limits of gravitational wave strain, are presented and discussed in Sections \ref{sec:results} and \ref{sec:discussion}, respectively and restated in Sec. \ref{sec:conclusion}.

\section{Search Method}\label{sec:sideband}

For a gravitational wave source in a binary system, the frequency of
the signal is Doppler modulated by the orbital motion of the
source with respect to the Earth ~\cite{RansomEA_2003,MW2007,SammutEA_2014}. The semi-coherent sideband search method involves the incoherent summation of frequency
modulated sidebands of the coherent \ftext-statistic ~\cite{JKS1998,MW2007,SammutEA_2014}. 

The first step in the sideband search is to calculate the coherent
\fstext{} as a function of frequency, assuming only a fixed sky
position. Knowing the sky position, one can account for the phase
evolution due to the motion of the detector. For sources in binary
systems, the orbital motion splits the signal contribution to the
\fstext{} into approximately $M=2m+1$ sidebands separated by $1/P$ in
frequency, where $m=\text{ceiling}(2\pi \fo \asini)$ \footnote{The $\text{ceiling}()$ function rounds up to the nearest integer.}, $\fo$ is the
intrinsic gravitational wave frequency, $\asini$ is the light travel
time across the semi-major axis of the orbit, and $P$ is the orbital
period. Knowledge of $P$ and $\asini$, allows us to construct an
\fstext{} sideband template.

The second stage of the sideband pipeline is the calculation of the
\cstext, where we convolve the sideband template with the coherent
\fstext{}. The result is an incoherent sum of the signal power at each
of the potential sidebands as a function of intrinsic gravitational
wave frequency. For our template we use a flat comb function with
equal amplitude teeth (see Fig. 1 of \cite{SammutEA_2014}), and hence, for a discrete frequency bin $f_k$, the template is given by
\begin{equation}\label{eq:template}
 \mathcal{T}(f_k) = \sum_{j=-m'}^{m'}\delta_{k\,l_{[j]}},
\end{equation}
where $m'=\text{ceiling}(2\pi f' a')$ depends on search frequency
$f'$ and the semi-major axis $a'$ used to construct the template (see
Sec.~\ref{subsec:asini})~\footnote{The frequency $f'$ is the central frequency of the
  search sub-band. Although the pipeline is designed for a wide-band
  search, the total band should be broken up into sub-bands narrow
  enough that the number of sidebands in the template, $m'(f')$, does
  not change significantly from the lower to the upper edge of the
  sub-band.}. The index $l_{[j]}$ of the Kronecker
delta-function is defined as
\begin{equation}\label{eq:findex}
  l_{[j]}\equiv\texttt{round}\left(\frac{j}{P' \Delta_f}\right),
\end{equation}
for a frequency bin width $\Delta f$, where round() returns the
closest integer, and $P'$ denotes our best guess at the orbital
period. The following convolution then yields the
\ctext-statistic,
\begin{eqnarray}
  \mathcal{C}(f_k) &=&\left(2\fstat\ast\mathcal{T}\right)(f_k) \label{eq:cstat_comb_conv} \\ 
     &=& \sum_{j=-m'}^{m'}2\fstat(f_{k-l_{[j]}}),\label{eq:cstat_comb}
\end{eqnarray}
where the form of $2\fstat$ is described in
\cite{JKS1998,Prix_2007,SammutEA_2014}. We therefore obtain this final
statistic as a function of frequency evaluated at the same discrete
frequency bins $f_k$ on which the input \fstext{} is computed.

\section{Parameter space}\label{sec:params}

\ac{ScoX1} is the brightest \ac{LMXB}, and the first to be discovered in
1962~\cite{GiacconiEA_1962}, located 2.8 kpc
away~\cite{BradshawEA_1999}, in the constellation Scorpius. Source
parameters inferred from a variety of electromagnetic measurements are
displayed in Table~\ref{tab:ScoX1_params}. Assuming that the
gravitational radiation and accretion torques balance, we obtain an
indirect upper limit on the gravitational wave strain amplitude for
Sco X-1 as a function of $\fspin$~\footnote{For a mass quadrupole
  $f_\subtext{GW} = 2\fspin$, while for a current quadrupole
  $f_\subtext{GW} =4\fspin/3$.}. Assuming fiducial values for the mass $\Mstar=1.4\Msol$, radius $R=10$ km
\cite{Steeghs_Casares_2002}, and moment of inertia $I=10^{38}$ kg m${}^2$ \cite{BradshawEA_1999} gives
\begin{equation}
 \ho^\text{EQ} \approx 3.5 \times 10^{-26} \left(\frac{300
\textrm{Hz}}{\fspin}\right)^{1/2} \label{eq:ScoX1_h0}.
\end{equation}
Equation~\ref{eq:ScoX1_h0} assumes that all the angular momentum
due to accretion is transferred to the star and converted into gravitational waves, providing an
upper limit on the gravitational wave strain ~\footnote{The torque-balance
  argument implies no angular momentum is lost from the system through
  the radio jets for example.}.

\begin{table}
  \caption{\label{tab:ScoX1_params}Sco X-1 observed parameters}
\begin{ruledtabular}
\begin{tabular}{lccclr}
Parameter (Name and Symbol) && && Value [Reference] & \\ \hline
X-ray flux && $\Fx$ && $4 \times 10^{-7}$ erg cm${}^{-2}$ s${}^{-1}$ & \cite{WattsEA_2008} \\ 
Distance && $D$ && $2.8 \pm 0.3$ kpc & \cite{BradshawEA_1999}\\ 
Right ascension && $\alpha$ && 16h 19m 55.0850s & \cite{BradshawEA_1999}\\ 
Declination && $\delta$ && ${-15}^\circ$ 38' 24.9" & \cite{BradshawEA_1999}\\ 
Sky Position angular resolution && $\Delta\beta$ && 0.3 mas & \cite{BradshawEA_1999}\\
Proper motion && $\mu$ && 14.1 mas yr${}^{-1}$ & \cite{BradshawEA_1999} \\
Orbital period && $P$ && $68023.70496 \pm 0.0432$ s &
\cite{GallowayEA_2014}\\
Projected semi-major axis && $\asini$ && 1.44 $\pm 0.18$ s &
\cite{Steeghs_Casares_2002}\\ 
Polarization angle && $\psi$  && $234 \pm 3\dg$ &
\cite{FomalontEA_2001b} \\
Inclination angle && $\iota$ && $44 \pm 6\dg$ &
\cite{FomalontEA_2001b}
\end{tabular}
\end{ruledtabular}
\end{table}

Optical observations of \ac{ScoX1} have accurately determined its sky
position and orbital period and, less accurately, the semi-major
axis~\cite{GottliebEA_1975,BradshawEA_1999,Steeghs_Casares_2002,GallowayEA_2014}. The
rotation period remains unknown, since no X-ray pulsations or bursts
have been detected. Although twin kHz \acp{QPO} have been observed in the contiuous X-ray flux with separations in the range $240-310$ Hz, there is no consistent and validated method that supports a relationship between the \ac{QPO} frequencys and the spin frequency of the neutron star (see~\cite{Watts_2012} for a review). We therefore assume the spin period is unknown and search over a range of $\fspin$. We also assume a circular orbit, which is expected by the time mass transfer occurs in \ac{LMXB} systems. In general, orbital eccentricity causes a redistribution of signal
power amongst the existing circular orbit sidebands and will cause
negligible leakage of signal power into additional sidebands at the
boundaries of the sideband structure. Orbital eccentricity also has
the effect of modifying the phase of each sideband. However, the standard sideband search is
insensitive to the phase of individual sidebands.

This section defines the parameter space of the sideband search,
quantifying the accuracy with which each parameter is and/or needs to
be known. The parameters and their uncertainties are summarised in
Table~\ref{tab:search_params}. 
\begin{table}
  \caption{\label{tab:search_params}Derived Sideband search
    parameters}
\begin{ruledtabular}
\begin{tabular}{lll}
Parameter & Symbol & Value   \\ \hline
Spin limited observation time & $\Ts^\text{spin}$ & 13 days \\ 
Period limited observation time \footnote{at $\fo=1$ kHz}& $\Ts^\text{Porb}$ & 50 days \\
Maximum sky position error& $\Delta\beta^\text{max}$ & 300 mas \\ 
Maximum proper motion \footnotemark[\value{mpfootnote}]\footnote{for $\Ts=10$days}& $\mu_\beta^\text{max}$ & 3000 mas yr${}^{-1}$ \\ 
Neutron star inclination \footnote{from EM observations} & $\iota_\subtext{EM}$ & $44 \dg \pm 6\dg$ \\ 
\hspace{1.5cm}  - EM independent & $\cos\iota$ & $[-1,\; 1]$ \\ 
Gravitational wave polaristion \footnotemark[\value{mpfootnote}] & $\psi_\subtext{EM}$ & $234 \dg \pm 4\dg$ \\ 
\hspace{1.5cm}  - EM independent & $\psi$ & $[-\tfrac{\pi}{4}, \; \tfrac{\pi}{4}]$  \\ 
\end{tabular}
\end{ruledtabular}
\end{table}

\subsection{Spin frequency} \label{sec:spinfreq}

The (unknown) neutron star spin period is likely to fluctuate due to
variations in the accretion rate $\dot{M}$. The coherent observation
time span $\Ts$ determines the size of the frequency bins in the
calculation of the \ftext-statistic, along with an over-resolution
factor $r$ defined such that a frequency bin is $1/(r\Ts)$ Hz wide. To
avoid sensitivity loss due to the signal wandering outside an individual frequency bin, we restrict the coherent observation time to less
than the spin limited observation time $\Ts^\text{spin}$ so that the signal is approximately monochromatic. Conservatively, assuming the deviation of the accretion torque from the mean flips sign randomly on the timescale $t_s {\sim} $ days~\cite{UBC_2000}, $\fspin$
experiences a random walk which would stay within a Fourier frequency
bin width for observation times less than $\Ts^\text{spin}$ given by
\begin{eqnarray} \label{eq:Ts_spin}
 \Ts^\text{spin} = \left(\frac{2\pi I}{r N_a}\right)^{2/3}\left(\frac{1}{t_s}\right)^{1/3},
\end{eqnarray} 
where $I=\tfrac{2}{5}\Mstar R^2$ is the moment of inertia of a neutron star
with mass $\Mstar$ and radius $R$, $N_a=\dot{\Mstar}(G\Mstar R)^{1/2}$ is the mean
accretion torque and $G$ is the gravitational constant.

For \ac{ScoX1}, with fiducial values for $\Mstar$, $R$, and $I$ as described earlier, and assuming $t_s = 1$ day (comparable to the timescale of fluctuations
in X-ray flux \cite{BildstenEA_1997}), $\Ts^\text{spin}=13$ days. We choose observation time span
$\Ts=10$ days to fit safely within this restriction.

\subsection{Orbital period}

The orbital period $\Porb$ sets the frequency spacing of the sidebands. Uncertainties in this parameter will
therefore translate to offsets in the spacing between the template and signal sidebands. The maximum coherent observation timespan
$\Ts^\text{Porb}$ allowed for use with a single template value of $\Porb$ is determined by the uncertainty $\Delta \Porb$ and can be expressed via
\begin{eqnarray}\label{eq:Ts_Porb}
\Ts^\text{Porb} \approx \frac{{\Porb}^2}{2\pi r\fo\asini |\Delta P|}, 
\end{eqnarray}
where a frequency bin in the \ftext-statistic has width $1/r\Ts$ as
explained above, and $\fo$ and $\asini$ are the intrinsic gravitational
wave frequency and light crossing time of the projected semi-major
axis respectively. For a \ac{ScoX1} search with $r=2$ at $\fo=1$ kHz,
one finds $\Ts^\text{Porb}=50$ days, longer than the maximum
duration allowed by spin wandering (i.e. $\Ts^\text{Porb} >
\Ts^\text{spin}$). Choosing $\fo=1$ kHz gives a conservative limit for $\Ts^\text{Porb}$ since lower frequencies will give higher values, and we only search up to 550 Hz. Thus we can safely assume the orbital period is known exactly for a search spanning $\Ts \leqslant$ 50 days. 

\subsection{Sky position and proper motion}

Knowledge of the source sky position is required to demodulate the
effects of detector motion with respect to the barycentre of the
source binary system (due to the Earth's diurnal and orbital motion) when
calculating the \ftext-statistic. We define an approximate worst-case
error in sky position $\Delta \beta^\text{max}$ as that which would
cause a maximum gravitational wave phase offset of 1 rad, giving us
\begin{equation} \label{eq:dbeta}
  |\Delta\beta^\text{max}|=\left(2\pi\fo R_\text{o}\right)^{-1},
\end{equation}
where $R_\text{o}$ is the Earth-Sun distance (1 AU).  Additionally,
the proper motion of the source also needs to be taken into
account. If the motion is large enough over the observation time it
will contribute to the phase error in the same way as the sky position
error. The worst case proper motion $\mu_{_\beta}^\text{max}$ can
therefore be determined similarly, viz.
\begin{equation} \label{eq:dmubeta}
 |\mu_{_\beta}^\text{max}| \leqslant (2\pi\fo R_\text{o} \Ts)^{-1} 
\end{equation}
For a 10-day observation at $\fo=1$ kHz, one finds $\Delta\beta=100$
mas and $\mu_{_\beta}^\text{max}= 3000 \text{ mas yr}^{-1}$.

The sky position of \ac{ScoX1} has been measured to within 0.3 mas,
with a proper motion of 14.1 mas
yr${}^{-1}$~\cite{BradshawEA_1999}. These are well within the allowed
constraints, validating the approximation that the sky position can be
assumed known and fixed within our analysis.

\subsection{Semi-major axis} \label{subsec:asini}

The semi-major axis determines the number of sidebands in the search
template. Its uncertainty affects the sensitivity of the search
independently of the observation time. To avoid the template width being underestimated, we construct a template using a semi-major axis $a'$ given by the (best guess) observed value $a$ and its uncertainty $\Delta a$ such that
\begin{equation}
 a' = a + \Delta a, 
\end{equation}
thus minimising signal losses. For a justification of this choice for $a'$, see Section IV. D. in \cite{SammutEA_2014}.

\subsection{Inclination and polarisation angles} \label{subsec:nuisance}

The inclination angle $\iota$ of the neutron star is the angle the spin axis makes
with respect to the line of sight. Without any observational prior we
would assume that the orientation of the spin axis is drawn from an isotropic distribution, and therefore $\cos\iota$ comes from a uniform distribution within the range $[-1,1]$. The polarisation angle $\psi$ describes the orientation of
the gravitational wave polarisation axis with respect to the
equatorial coordinate system, and can be determined from the position
angle of the spin axis, projected on the sky. Again, with no
observational prior we assume that $\psi$ comes from a uniform distribution
within the range $[0,2\pi]$.

The orientation angles $\iota$ and $\psi$ affect both the
amplitude and phase of the incident gravitational wave.  The phase contribution can be
treated separately from the binary phase and the uncertainty in both
$\iota$ and $\psi$ are analytically maximised within the construction
of the \ftext-statistic. However, electromagnetic observations can be
used to constrain the prior distributions on $\iota$ and $\psi$.  This
information can be used to improve search sensitivity in
post-processing when assessing the response of the pipeline to signals
with parameters drawn from these prior distributions. 

In this paper, we consider two scenarios for $\iota$ and $\psi$: (i)
uniform distributions within the previously defined ranges; and (ii) prior distributions based on
values and uncertainties obtained from electromagnetic
observations. From observations of the radio jets from
\ac{ScoX1}~\cite{FomalontEA_2001b} we can take $\iota = 44\dg \pm
6\dg$, assuming the rotation axis of the neutron star is perpendicular
to the accretion disk. The same observations yield a position angle of
the radio jets of $54\pm 3\dg$. Again, assuming alignment of the spin
and disk normal, the position angle is directly related to the
gravitational wave polaristion angle with a phase shift of $180\dg$,
such that $\psi = 234\pm 3\dg$.  For these observationally motivated
priors we adopt Gaussian distributions, with mean and variance given by the observed values and their errors, respectively, as quoted above.

\section{Implementation}\label{sec:implement}

\subsection{Data selection}\label{subsec:data}


\ac{LIGO}'s fifth science run (S5) took
place between November 4, 2005 and October 1, 2007. During this period
the three LIGO detectors (L1 in Livingston, LA; H1 and H2 collocated
in Hanford, WA) achieved approximately one year of triple coincidence
observation, operating near their design
sensitivity~\cite{LIGO_S5cal_Abadie2010}. The amplitude spectral
density of the strain noise of the two 4-km detectors (H1 and L1) was
a minimum $3\times10^{-23}\text{Hz}^{-1/2}$ at 140 Hz and $\lesssim
5\times 10^{-23}\text{ Hz}^{-1/2}$ over the 100-300 Hz band.

\begin{figure}
\includegraphics[width=\columnwidth]{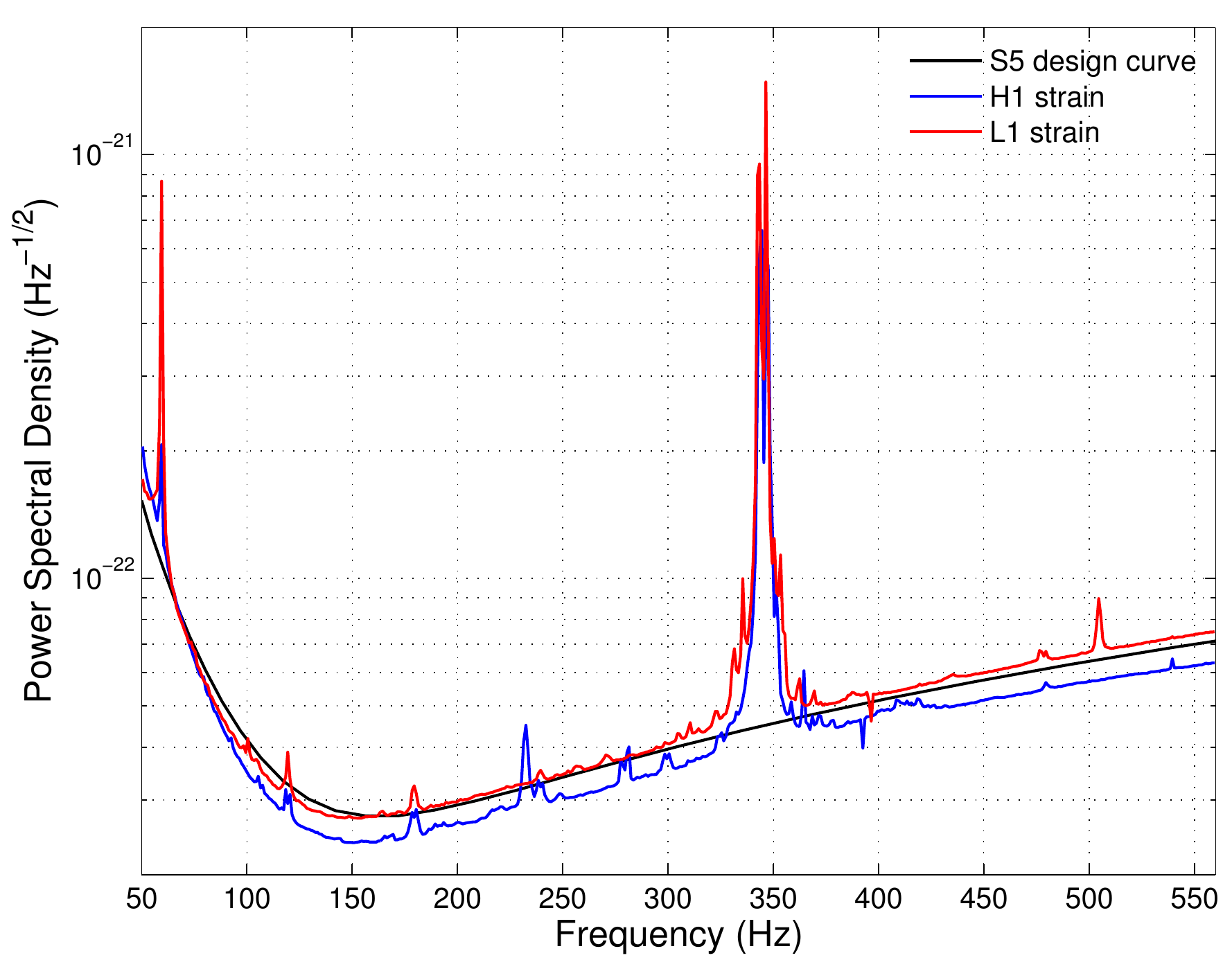}
\caption{\label{fig:PSD_compare}(colour online) \ac{LIGO} S5 strain sensitivity curve (black) compared to power spectral density of both H1 (blue, lower) and L1 (red, upper) detectors during the selected 10 day data stretch, which ran from 21--31 August 2007 (GPS time 871760852--872626054).}
\end{figure}

Unkown effects of accretion on the rotation period of the neutron star (spin wandering) restricts the sideband analysis to a 10-day coherent observation
time span (Section~\ref{sec:spinfreq}). A 10-day data-stretch was
selected from S5 as follows~\cite{LIGO_CasA_2010}. A figure of merit,
proportional to the \ac{SNR} and defined by
$\sum_{k,f}[S_h(f)]_k^{-1}$, where $[S_h(f)]_k$ is the strain noise
power spectral density at frequency $f$ in the $k^{\text{th}}$
\ac{SFT}, was assigned to each rolling 10-day stretch. The highest
value of this quantity over the 100--300 Hz band (the region of
greatest detector sensitivity) was achieved in the interval August
21--31, 2007 (GPS times 871760852--872626054) with duty factors of 91
\% in H1 and 72\% in L1. This data stretch was selected for the
search. We search a 500 Hz band, ranging from 50-550 Hz, chosen to include the most sensitive region of the detector. The power spectral density for this stretch of data is shown in Fig. \ref{fig:PSD_compare}. The most prominent peaks in the noise spectrum are due to power line harmonics at 60 Hz and thermally excited violin modes from 330--350 Hz caused by the mirror suspension wires in the interferometer \cite{speclines}.

Science data (data that excludes detector down-time and times flagged with
poor data quality) are calibrated to produce a strain time
series $h(t)$, which is then broken up into shorter segments of equal
length. Some data are discarded, as not every continuous section of
$h(t)$ covers an integer multiple of segments. The segments are high-pass filtered above 40 Hz and Fourier transformed to form
\acp{SFT}. For this search, 1800 sec \acp{SFT} are fed into the
\fstext{} stage of the pipeline.

\subsection{Pipeline}\label{subsec:pipeline}  
  
\begin{figure}
\begin{tikzpicture}[node distance = 1.5cm, auto]
    \node [terminal] (start) {Select target and data};
    \node [process, below of=start, node distance=1.5cm] (cfs_iso) {compute \fstext};
    \node [data, right of=cfs_iso, node distance=3.5cm] (skypos) {$\alpha$, $\delta$};
    \node [data, below of=cfs_iso, node distance=1.5cm] (fstats) {\ftext-statistic output \hspace{.8cm}    (zero threshold)};
    \node [process, below of=fstats, node distance=1.5cm] (comb){compute \cstext};
    \node [data, right of=comb, node distance=3.5cm] (inputs) {$\Porb$, $\asini$};
    \node [data, below of=comb, node distance=1.5cm] (cstats)
{\ctext-statistic output};
    \node [decision, below of=cstats, node distance=1.8cm] (threshold) {$\cstat>\cstar$?};
    \node [process, below of=threshold, node distance=2cm] (bayes)
{Bayesian \hspace{0.25cm} upper limits};
    \node [decision, right of=bayes, node distance=3cm] (noise) {Artifact?};
    \node [terminal, below of=bayes, node distance=2cm] (uls) {Upper Limits};
    \node [terminal, right of=uls, node distance=3cm] (signal) {Detection};
    \path [line] (start) -- (cfs_iso); 
    \path [line] (cfs_iso) -- (fstats);
    \path [line] (skypos) -- (cfs_iso);
    \path [line] (fstats) -- (comb);
    \path [line] (inputs) -- (comb);
    \path [line] (comb) -- (cstats);
    \path [line] (cstats) -- (threshold);
    \path [line] (threshold) -- node [right] {No} (bayes);
    \path [line] (threshold) -| node [near start, above] {Yes} (noise);
    \path [line] (noise) -- node [right] {No} (signal);
    \path [line] (noise) -- node [above]  {Yes} (bayes);
    \path [line] (bayes) -- (uls);
\end{tikzpicture}
\caption{\label{fig:pipeline}Flowchart of the search
  pipeline. After data selection, the \fstext for an isolated source is calculated in the compute \fstext stage with the source sky position ($\alpha$, $\delta$) as input. The output of this is then passed to the sideband search in the compute \cstext stage, with the binary parameters ($\Porb$ and $\asini$) as input, which returns a \cstext. When $\cstat$ is greater than the threshold $\cstar$, the candidate is investigated as a potential signal. If no candidates survive follow-up, upper limits are presented.}
\end{figure}
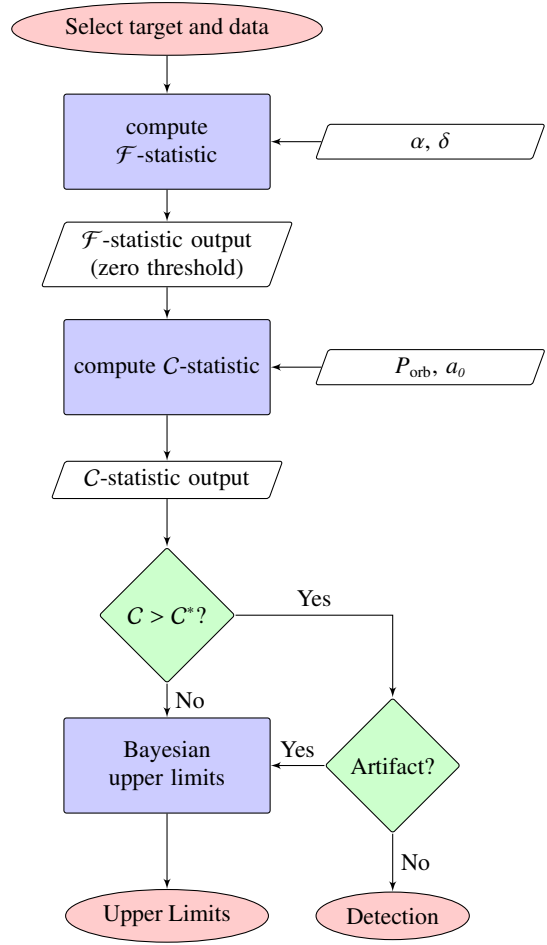

A flowchart of the multi-stage sideband pipeline is depicted in
Fig.~\ref{fig:pipeline}. After data selection, the first stage of the pipeline is the
computation of the \fstext{} ~\cite{Prix_2011_cfsv2,lal}~\footnote{Note that there is no thresholding applied to the \fstext.}. For
the sideband search only the sky position is required at the \fstext{}
stage, where the matched filter models an isolated source. 

The outputs of the \fstext{} analysis are values of $2\fstat$ for
each frequency bin from which the sideband algorithm then calculates the
\cstext{}~\cite{SammutEA_2014,lal}. The algorithm takes values of the
\fstext{} as input data and values of $\Porb$ and $\asini$ as input
parameters, and outputs a \cstext{} for every frequency bin in the
search range (as per Eqs. \ref{eq:cstat_comb_conv} and
\ref{eq:cstat_comb}).

The extent of the sideband template, Eq.~\ref{eq:template}, changes as a
function of the search frequency $f'$ since the number of sidebands in
the template scales as $M\propto f'$. We therefore divide the 500 Hz search band into smaller sub-bands over which we can use a single template. The sub-bands must be narrow enough, so that $f'$ and hence $M$ do not change significantly from
the lower to the upper edges of the sub-band, and wide enough to
contain the entire sideband pattern for each value of $f'$. It is preferable to generate \fstext{} data files matching these sub-bands,
so that the search algorithm can call specific \fstext{} data
files for each template, as opposed to each call being directed to the
same large data file. However, the \fstext sub-bands need to be half a sideband width (or $2\pi f'a'/P'$ Hz) wider on each end than the \cstext sub-bands in order to calculate the \cstext at the outer edges. For a \ac{ScoX1} directed search,
single-Hz bands are convenient; for example, even up at $f'=1000$ Hz, the
template width $4\pi f' a/P$ is still less than 0.25 Hz.

The output of the \cstext{} is compared with a threshold value
$\cstar$ chosen according to a desired false alarm rate (see Sec. \ref{subsec:threshold} below). Any frequency
bins returning $\cstat>\cstar$ are designated as candidate events and
are investigated to determine whether they can be attributed to non-astrophysical origins, due to noise or detector artifacts, or to an
astrophysical signal. The former are vetoed and if no candidates
above $\cstar$ survive, upper limits are computed (see Sec. \ref{subsec:veto} for more information on the veto procedure).

%
  
\subsection{Detection Threshold}\label{subsec:threshold}

To define the threshold value $\cstar$ for a \emph{single} trial we first
relate it to the false alarm probability $\Pa$, i.e. the probability
that noise alone would generate a value greater than this threshold.
This is given by
\begin{eqnarray}
 \Pa &=& p(\cstat>\cstar|\text{no signal}) \nonumber \\
     &=& 1 - F(\cstar,4M),\label{eq:Pa}
\end{eqnarray}
where $F(x,k)$ denotes the cumulative distribution function of a
$\chi^2_k$ distribution evaluated at $x$ \footnote{The \cstext is a $\chi^2$ distributed variable with $4M$ degrees of freedom because it is the sum over $M$ sidebands of the \fstext, which is in turn $\chi^2$ distributed with 4 degrees of freedom}.

In the case of $N$ statistically independent trials, the false alarm
probability is given by
\begin{eqnarray}
 \PaN &=& 1-(1-\Pa)^N \nonumber \\
         &=& 1-\left[F(\cstar,4M)\right]^N.
\end{eqnarray}
This can be solved for the detection threshold $\cstar_N$ in the case
of $N$ trials, giving
\begin{align}
\cstar_N &= F^{-1}([1-P_{\text{a}|N}]^{1/N}, 4M), \label{eq:cstar_N}
\end{align}
where $F^{-1}$ is the inverse (not the reciprocal) of the function
$F$.


The search yields a different \cstext for each frequency bin in the
search range. If the \cstext values are uncorrelated, we can equate
the number of independent trials with the number of independent
frequency bins ($\propto T$ for each Hz band). However, due to the
comb structure of the signal and template, frequencies separated by an
integer number of frequency-modulated sideband spacings become
correlated, since each of these values are constructed from sums
of \fstext values containing many common values. The pattern of $M$
sidebands separated by $1/P$ Hz spans $M/P$ Hz, meaning there are
$P/M$ sideband patterns per unit frequency. Hence, as an
approximation, it can be assumed that within a single comb
template there are $T/P$ independent \cstext results. The number of
statistically independent trials per unit Hz is therefore given by the
number of independent results in one sideband multiplied by the number
of sidebands per unit frequency, i.e.
\begin{align}
 N &\approx\frac{T}{M}.
\end{align}
This is a reduction by a factor $M$ in the number of statistically
independent \cstext values as compared to the \fstext.

Using this more realistic value of $N$ provides a better analytical
prediction of the detection threshold for a given $\Pa$, which we can
apply to each frequency band in our search. However, a precise determination of significance (taking into account correlations between different C-statistics among other effects) requires a numerical investigation. To factor this in,
we use \ac{MC} simulations to estimate an approximate performance (or
loss) factor, denoted $\kappa$. This value is estimated for a handful
of 1 Hz wide frequency bands and then applied across the entire search
band. For a specific 1 Hz frequency band, the complete search is
repeated 100 times, each with a different realisation of Gaussian
noise. The maximum \ctext obtained from each run is returned, and this
distribution of values allows us to estimate the value $\cstar_{MC}$
corresponding to a multi-trial false alarm probability
$P_{\text{a}|N}=1\%$. For each trial frequency band, we can then
estimate $\kappa$ as
\begin{equation}\label{eq:kappa}
 \kappa =  \frac{\cstar_{MC} - 4M}{\cstar_N - 4M} - 1,
\end{equation}
which can be interpreted as the fractional deviation in \ctext using
the number of degrees of freedom (the expected mean) as the point of
reference.  We assume that $\kappa$ is approximately independent of
frequency (supported by Table~\ref{tab:kappa}) and hence use a single
$\kappa$ value to represent the entire band. We incorporate this by
defining an updated threshold
\begin{equation}\label{eq:cstar_kappa}
 \cstar_\kappa = \cstar_N(1+\kappa) - 4M\kappa
\end{equation}
which now accounts for approximations in the analysis pipeline. The \ac{MC} procedure was performed for 1-Hz frequency bands starting at 55, 255, and 555 Hz. Using the values returned from these bands, we take a value of $\kappa=0.3$. Table \ref{tab:kappa}
lists the values of $4M$, $\cstar_N$, $\cstar_{MC}$ and $\kappa$ associated with each of these bands.
\begin{table}
\caption{\label{tab:kappa} Performance factor obtained from \ac{MC} simulations at
three different 1-Hz sub-bands. The starting frequency of the sub-band is listed in the first column.
The expected C-statistic value in Gaussian noise $4M$, theoretical threshold $\cstar_N$, and the threshold obtained from the MC simulations $\cstar_{MC}$ are listed in the second, third and fourth columns, respectively. Performance factor $\kappa$ is listed in the last column.}
\begin{ruledtabular}
\begin{tabular}{ccccc}
sub-band (Hz)	& $4M$  &  $\cstar_N$   & $\cstar_{MC}$	& $\kappa$  \\ \hline 
55  		& 4028  &  4410		& 4520  	& 0.28    \\
255		& 18500 &  19254 	& 19476 	& 0.29    \\ 
555 		& 40212 &  41264 	& 41578 	& 0.30    \\
\end{tabular}
\end{ruledtabular}
\end{table}
%

\subsection{Upper limit calculation}

If no detection candidates are identified, we define an
upper limit on the gravitational wave strain $\ho$ as the value $\hUL$
such that a predefined fraction $\pUL$ of the marginalised posterior
probability distribution $p(\ho|\cstat)$ lies between 0 and $\hUL$.
This value is obtained numerically for each \cstext by solving
\begin{equation}
 \pUL = \int\limits_{0}^{\hUL}p(\ho|\cstat) \,\, d\ho, \label{eq:h0_UL}
\end{equation}
with
\begin{eqnarray}\label{eq:h0_post}
  p(\ho|\cstat)\!&\propto&\!\int\limits_{-\infty}^{\infty}\!dP
\!\int\limits_{-\infty}^{\infty}\!da\!\int\limits_{0}^{2\pi}\!d\psi
\!\int\limits_{-1}^{1}\!d\cos\iota~p(\cstat|\bm{\theta})\mathcal{N}(a,\Delta{a})\mathcal{N}(P,\Delta{P}), \nonumber \\
\end{eqnarray}
and where $\mathcal{N}(\mu,\sigma)$ denotes a Gaussian (normal) distribution with mean $\mu$ and standard
deviation $\sigma$. The likelihood function $p(\cstat|\bm{\theta})$ is the \ac{pdf} of a non-central $\chi^2_{4M}(\lambda(\bf{\theta}))$ distribution given by
\begin{equation}\label{eq:likelihood}
 p(\cstat|\bm{\theta}) = \frac{1}{2}
\exp\left(-\frac{1}{2}\left(\cstat+\lambda\left(\bm{\theta}\right)\right)\right)
\left(\frac{\cstat}{\lambda\left(\bm{\theta}\right)}\right)^{M-\frac{1}{2}}
\textrm{I}_{2M-1}\left(
\sqrt{\cstat \lambda\left(\bm{\theta}\right)}\right),
\end{equation}
where $\textrm{I}_{\nu}(z)$ is the modified Bessel function of the first kind with order $\nu$ and argument $z$. The non-centrality parameter $\lambda(\bm{\theta)}$ is proportional to the optimal \ac{SNR} (see Eq. 64 of
\cite{SammutEA_2014}), and is a function of $\psi$, $\cos \iota$ and the mismatch in the template caused by $\Delta P$ and $\Delta a$. See Section \ref{subsec:nuisance} for a description on the priors selected for $\cos \iota$ and $\psi$.


It is common practice in continuous-wave searches to compute
frequentist upper limits using computationally expensive Monte Carlo
simulations. The approach above allows an upper limit to be
computed efficiently for each \cstext, since $p(\ho|\cstat)$ is calculated analytically instead of numerically and is a
monotonic function of \ctext. We also note that for large parameter
space searches the multi-trial false alarm threshold corresponds to
relatively large \ac{SNR} and in this regime the Bayes and frequentist
upper limits have been shown to converge~\cite{RMP2011}.

\section{Results}\label{sec:results}

We perform the sideband search on 10 days of \ac{LIGO} S5 data
spanning 21--31 August 2007 (see Fig.~\ref{fig:PSD_compare} and
Sec.~\ref{subsec:data}). The search covers the band from 50 -- 550 Hz. A \cstext is generated for each of the
$2\times 10^6$ frequency bins in each 1-Hz sub-band. The maximum
\cstext from each sub-band is compared to the theoretical threshold
(Eq.~\ref{eq:cstar_kappa}). Any \cstext above the threshold is classed
as a detection candidate worthy of further investigation.

Without pre-processing (cleaning) of the data, non-Gaussian instrumental noise and instrumental
artifacts had to be considered as potential sources for candidates. A comprehensive list of
known noise lines for the S5 run, and their origins, can be found in
Appendix B of~\cite{LIGO_E@H_S5_2013}. Candidates in sub-bands contaminated by these lines have been automatically removed. The veto described in Sec.~\ref{subsec:veto} was then applied to the remaining candidates in order to eliminate
candidates that could not originate from an astrophysical signal represented by our model. This
veto stage is first applied as an automated process but each candidate
is also inspected manually as a verification step.  If all candidates
are found to be consistent with noise, no detection is claimed, and
upper limits are set on the gravitational wave strain tensor amplitude
$h_0$.

\subsection{Detection candidates} \label{subsec:candidates}

\begin{flushleft}
\begin{figure*}
\includegraphics[width=\textwidth]{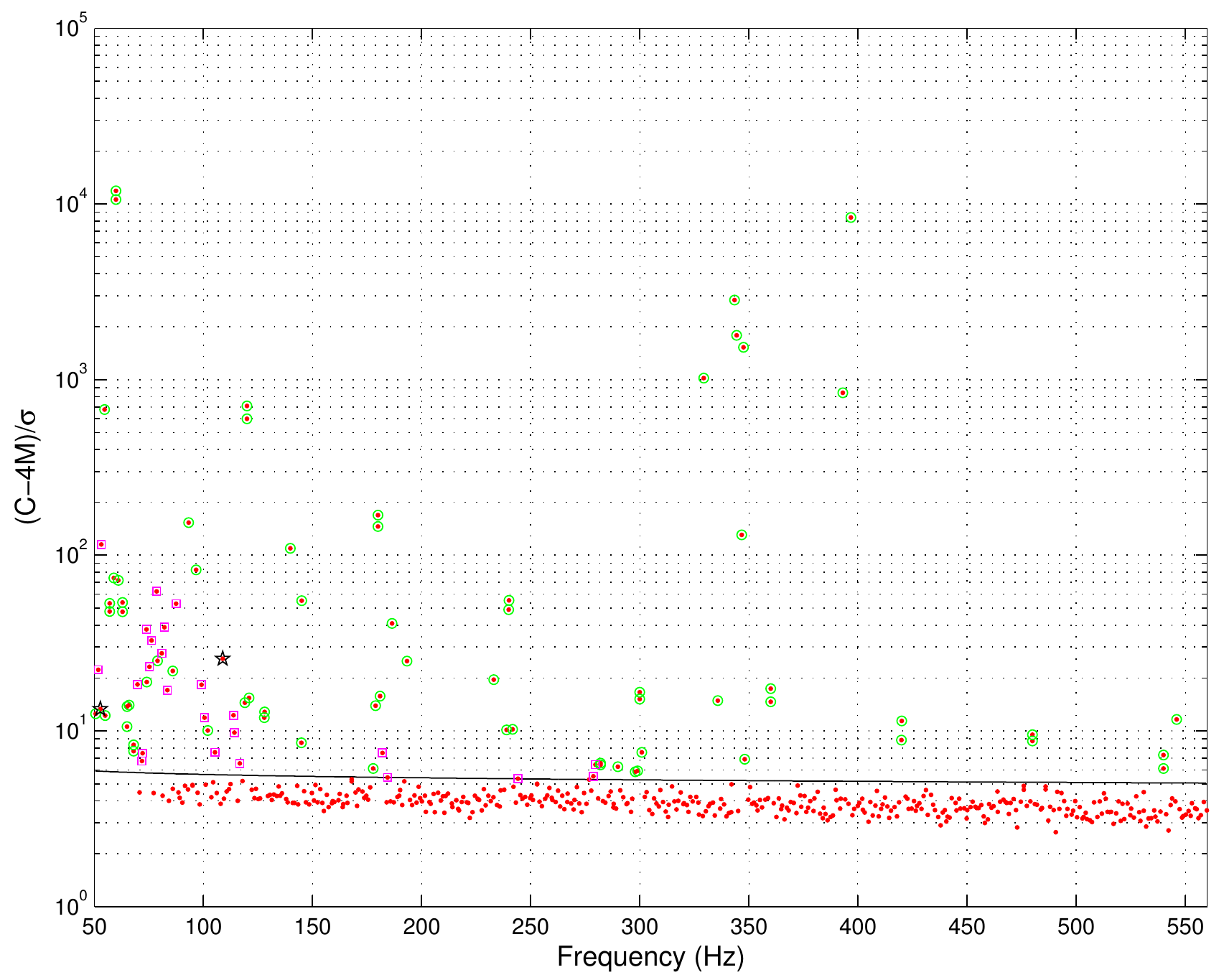}
\caption{\label{fig:Cstats}(colour online) Red dots indicate the maximum detection statistic for each Hz sub-band (reduced by the expected value $\text{E}[\cstat]=4M$ and normalised by the expected standard deviation $\sigma=\sqrt{8M}$) plotted as a function of frequency. The
  threshold value $\cstar_N$ for $N=T/M$ trials and $\PaN=1\%$ false
  alarm probability is shown for comparison (solid black
  curve). Points exceeding the threshold are marked by green circles
  if they coincide with a frequency band known to be contaminated by
  instrumental noise lines, black stars to indicate hardware-injected
  isolated pulsars, or pink squares to mark candidates requiring
  further investigation (follow-up).}
\end{figure*} 
\end{flushleft}

The maximum \cstext, $\cstat_\text{max}$, returned from each sub-band
is plotted in Fig.~\ref{fig:Cstats} as a function of frequency. The threshold
$\cstar_\kappa$ for $N=T/M$ trials and $\PaN=1\%$ false alarm probability is indicated by a solid black curve. Data points above
this line are classed as detection candidates. Candidates in sub-bands
contaminated by known instrumental noise are highlighted by green
circles and henceforth discarded. The two candidates highlighted by a
black star coincide with hardware-injected isolated pulsar signals
``5'' and ``3'' at $f= 52.808324$ and 108.85716 Hz respectively
(see Table III. in Section VI. of~\cite{LIGO_E@H_S5_2013} for more
details on isolated pulsar hardware injections).

The remaining candidates above the threshold are highlighted by pink
squares and merit follow-up. Their frequencies $f_\text{max}$ and
corresponding $\cstat_\text{max}$ and predicted threshold detection candidate $\cstar_\kappa$ are listed in
Table~\ref{tab:candidates}. The detection threshold with fixed
$\kappa$ varies with $f$ through the variation in the degrees of
freedom in the \ctext (see Sec.~\ref{subsec:threshold}).

\begin{table}
\caption{\label{tab:candidates} Maximum \cstext from each Hz sub-band
  exceeding the detection threshold $\cstar_N$ for $N$ trials after
  removing isolated pulsar injections and candidates in bands
  contaminated by known noise lines. The first column lists the
  frequency $f_{max}$ at which the maximum \cstext $\cstat_{max}$
  occurs. $\cstat_{max}$ and $\cstar_\kappa$ are listed in the second and third columns, respectively, for comparison.}
\begin{ruledtabular}
\begin{tabular}{rcc}
$f_{\text{max}}$ (Hz)&  $\cstat_{\text{max}} (\times 10^3)$ & 	$\cstar_\kappa (\times 10^3)$ \\ \hline 
51.785819 &   5.66 &   4.22 \\ 
53.258119 &   14.0 &   4.38 \\ 
69.753009 &   6.88 &    5.6 \\ 
71.879543 &   5.87 &   5.75 \\ 
72.124267 &   6.02 &   5.82 \\ 
73.978239 &   9.23 &   5.91 \\ 
75.307963 &   7.90 &   6.06 \\ 
76.186649 &   9.00 &   6.13 \\ 
78.560484 &   12.3 &   6.28 \\ 
80.898939 &   8.82 &   6.43 \\ 
82.105904 &   10.2 &   6.58 \\ 
83.585249 &   7.93 &   6.66 \\ 
87.519459 &   12.3 &   6.96 \\ 
99.113480 &   9.41 &   7.87 \\ 
100.543741 &   8.72 &   7.94 \\ 
105.277878 &   8.58 &   8.31 \\ 
113.764264 &   9.80 &   8.91 \\ 
114.267062 &   9.55 &   8.99 \\ 
116.686578 &   9.29 &   9.14 \\ 
182.150449 &   14.4 &   14.1 \\ 
184.392065 &   14.3 &   14.2 \\ 
244.181829 &   18.7 &   18.7 \\ 
278.712575 &   21.3 &   21.2 \\ 
279.738235 &   21.5 &   21.3 \\ 

\end{tabular}
\end{ruledtabular}
\end{table}
%

\subsection{Noise veto} \label{subsec:veto}

\begin{table}[h!]
  \caption{\label{tab:noise_veto} Candidates surviving the $4M$ veto. The table lists the start frequency of the 1-Hz sub-band containing the candidate, the expected \cstext value $4M$, the $\PaN=1\%$ threshold $\cstar_\kappa$, and the fraction of \cstexts below $4M$ and above $\cstar_\kappa$ in the range $|f-f_\text{max}| < M/P$ centred at the bin $f_{\text{max}}$ returning $\cstat_\text{max}$. The * marks the bands containing the candidates that survive the final, manual veto.}
\begin{ruledtabular}
\begin{tabular}{lllccl}
 $f_{\text{band}}$ (Hz) &  $4M$ &   $\cstar_\kappa$ & $\% < 4M$   &	$\% > \cstar_\kappa$ &  \\ \hline
69 & 5036 & 5596 & 15.3 & 52.5 & 	 \\ 
71 & 5180 & 5746 & 1.04 & 1.97 & 	 \\ 
105 & 7644 & 8314 & 1.17 & 4.02 & 	 \\ 
116 & 8436 & 9135 & 1.34 & 1.34 & 	 \\ 
184* & 13356 & 14208 & 27.0 & 0.0204 & 	 \\ 
244* & 17700 & 18662 & 33.2 & 0.00723 & 	 \\ 
278* & 20164 & 21182 & 14.7 & 0.0365 & 	 \\ 
279 & 20236 & 21255 & 4.71 & 4.5 & 	 \\ 
\end{tabular}
\end{ruledtabular}
\end{table}

Frequency bins coincident with signal sidebands generate \cstext
values drawn from a $\chi^2_{4M}(\lambda)$ distribution, while
frequency bins falling between these sidebands (the majority of
bins) should follow the noise distribution $\chi^2_{4M}(0)$. If noise
produces a spuriously loud \fstext in one bin, it then contributes
strongly to every \cstext in a sideband width centred on the spurious bin, a frequency range spanning ${\sim}
2M/P$, to the point where all the \cstexts may
exceed the expected mean $4M$ in this region. We can exploit this
property to design a veto against candidates occurring from noise
lines as follows: a candidate is vetoed as a potential astrophysical
signal if the fraction of bins with $\cstat < 4M$ in the range
$|f-f_\text{max}| < M/P$ is too low, where $f_\text{max}$ is the
frequency bin corresponding to $\cstat_\text{max}$. We set the
minimum bin fraction to zero so as not to discard a real signal strong enough
to make $\cstat>4M$ over a broad range. 

Applying the veto reduces the
number of candidate events from 24, shown in
Table~\ref{tab:candidates} to the eight listed in
Table~\ref{tab:noise_veto}. The eight candidates were inspected manually to identify if the features present are consistent with a signal (see Appendix \ref{ap:veto}). After manual inspection, three candidates remained, which could not be conclusively identified as a signal, but could still be expected from noise given the $1\%$ false alarm threshold set \footnote{The automated and manual veto stages were tested extensively on software injected signals and simulated Gaussian noise to ensure signals were not discounted accidentally.}. These final three candidates were contained in the 184, 244 and 278 Hz sub-bands and were followed up in two other 10-day stretches of S5 data.

\subsection{Candidate follow-up} \label{subsec:follow-up}

\begin{figure}
\includegraphics[width=\columnwidth]{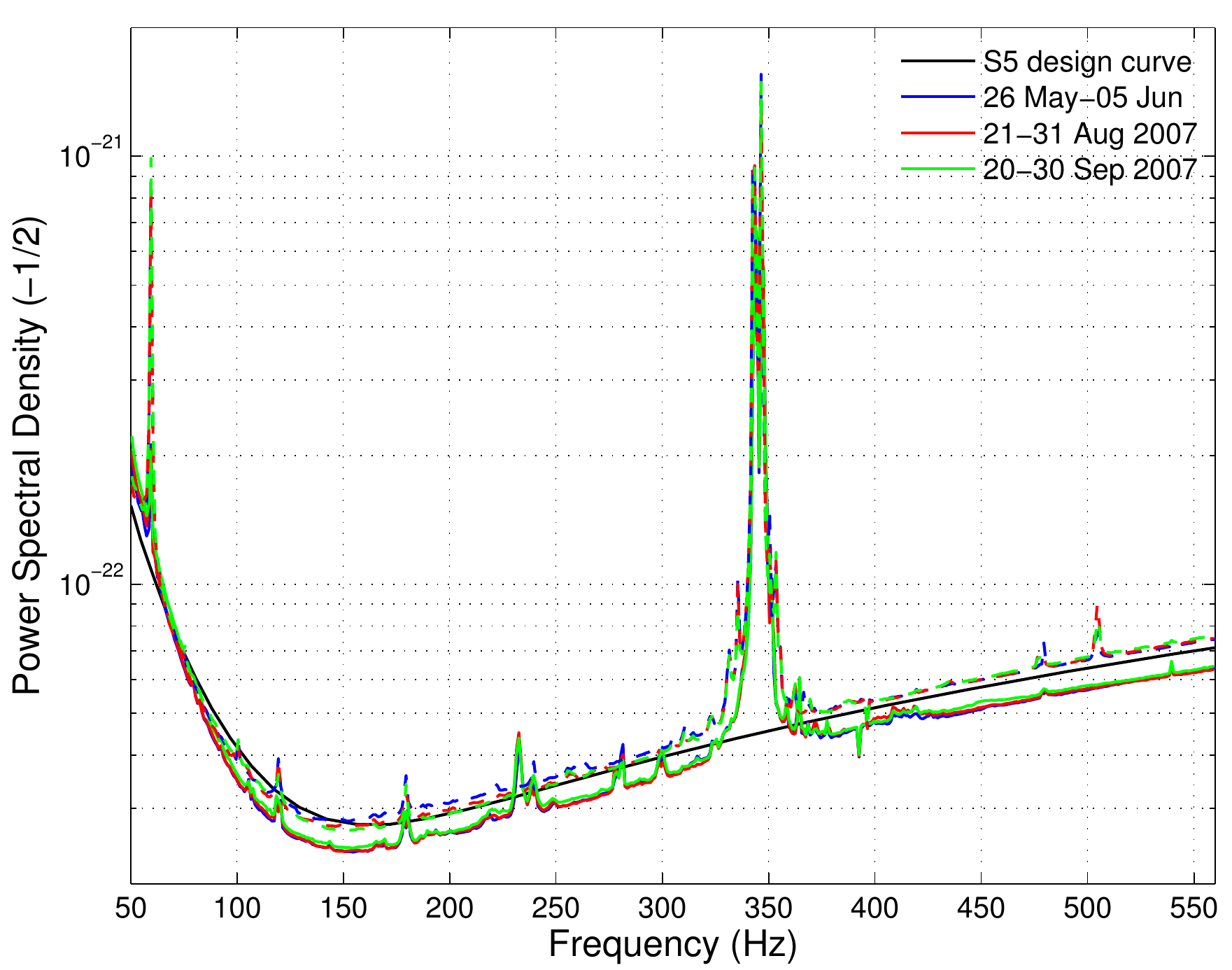}
\caption{\label{fig:PSD_compare_all}(colour online) \ac{LIGO} S5 strain sensitivity design curve (black) compared to
  power spectral density of both H1 (solid, lower) and L1 (dashed, upper) detectors
  during the selected 10 day data stretch (red), which ran from 21--31
  August 2007, and the other two stretches used for follow-up (26 May -- 05 Jun 2007 indicated in blue, and 20 -- 30 Sep 2007 in green).}
\end{figure}

The three remaining candidate bands were followed up by analysing two other 10-day stretches of S5 data of comparable sensitivity. A comparison of the noise spectral density of each of the 10-day stretches is displayed in Fig. \ref{fig:PSD_compare_all} and the results from each of the three bands are presented in Table \ref{tab:follow-up}. The bands did not produce significant candidates in the two follow-up searches, indicating they were noise events. This is indicated more robustly by the combined P-values for each candidate presented at the bottom of Table \ref{tab:follow-up}.

All three candidates lie at the low frequency end, in the neighbourhood
of known noise lines (green circles identify excluded points in
Fig. \ref{fig:Cstats}) and may be the result of noise-floor
fluctuations (caused by the non-stationarity of seismic noise, which dominates the noise-floor at low frequencies). Events such as these are expected to occur from noise in $1\%$ of cases, as defined by our false alarm threshold, and are consistent with the noise hypothesis. 

\begin{table}
\caption{\label{tab:follow-up} Results from candidate follow-up for each observation timespan at each Hz frequency band. The fractional percent above $\cstar_\kappa$ and below $4M$ are taken from the expected signal region indicated by the original candidate (which includes a sideband width centred at the candidate, plus the maximum effects of any spin wandering). The P-value is calculated for the maximum \cstext value in this region. A combined P-value for each candidate is displayed at the bottom of the table.}
\begin{ruledtabular}
\begin{tabular}{c|c|cccc}
Timespan      	&          & 184 Hz & 244 Hz & 278 Hz & \\ \hline
21--31 Aug   	& \% above $\cstar_\kappa$  & 0.02   & 0.01    & 0.04   & \\ 
2007		& \% below $4M$   & 27.02  & 33.22  & 17.72  & \\
(original)	& P-value       & $1.03\times 10^{-5}$ & $1.12\times 10^{-5}$ & $5.36\times 10^{-6}$  & \\ \hline
20 -- 30 Sep 	& \% above $\cstar_\kappa$  & 0.00   & 0.00   & 0.00  & \\
2007		& \% below $4M$   &  32.46 & 46.69  & 33.54 & \\
(follow-up)   & P-value  	& 0.92   & 0.27   & 0.36  & \\ \hline
26 May -- 05 Jun& \% above $\cstar_\kappa$  & 0.00 	 & 0.00   & 0.00  & \\
2007 		& \% below $4M$   & 41.44  & 49.27  & 63.38 & \\
(follow-up)   & P-value       & 0.50   & 0.47   & 0.15  & \\ \hline\hline
Combined  	& P-value  	& 0.99 	 & 0.75   & 0.60  &
\end{tabular}
\end{ruledtabular}
\end{table}
%

\subsection{Upper limits}

\begin{figure*}
\subfigure[\label{subfig:ULs}]{
\includegraphics[width=0.46\textwidth]{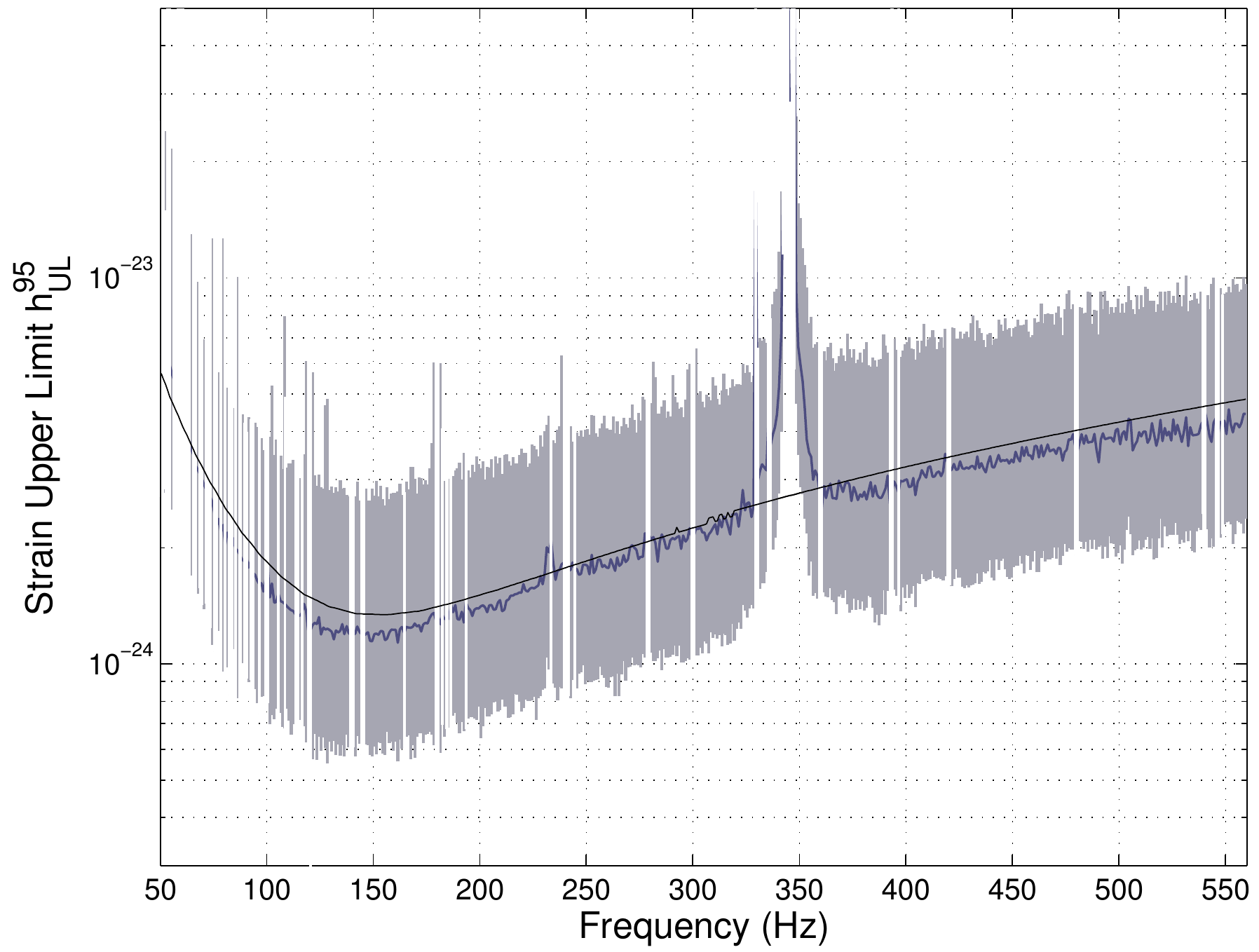}}
\subfigure[\label{subfig:ULs_dir}]{
\includegraphics[width=.46\textwidth]{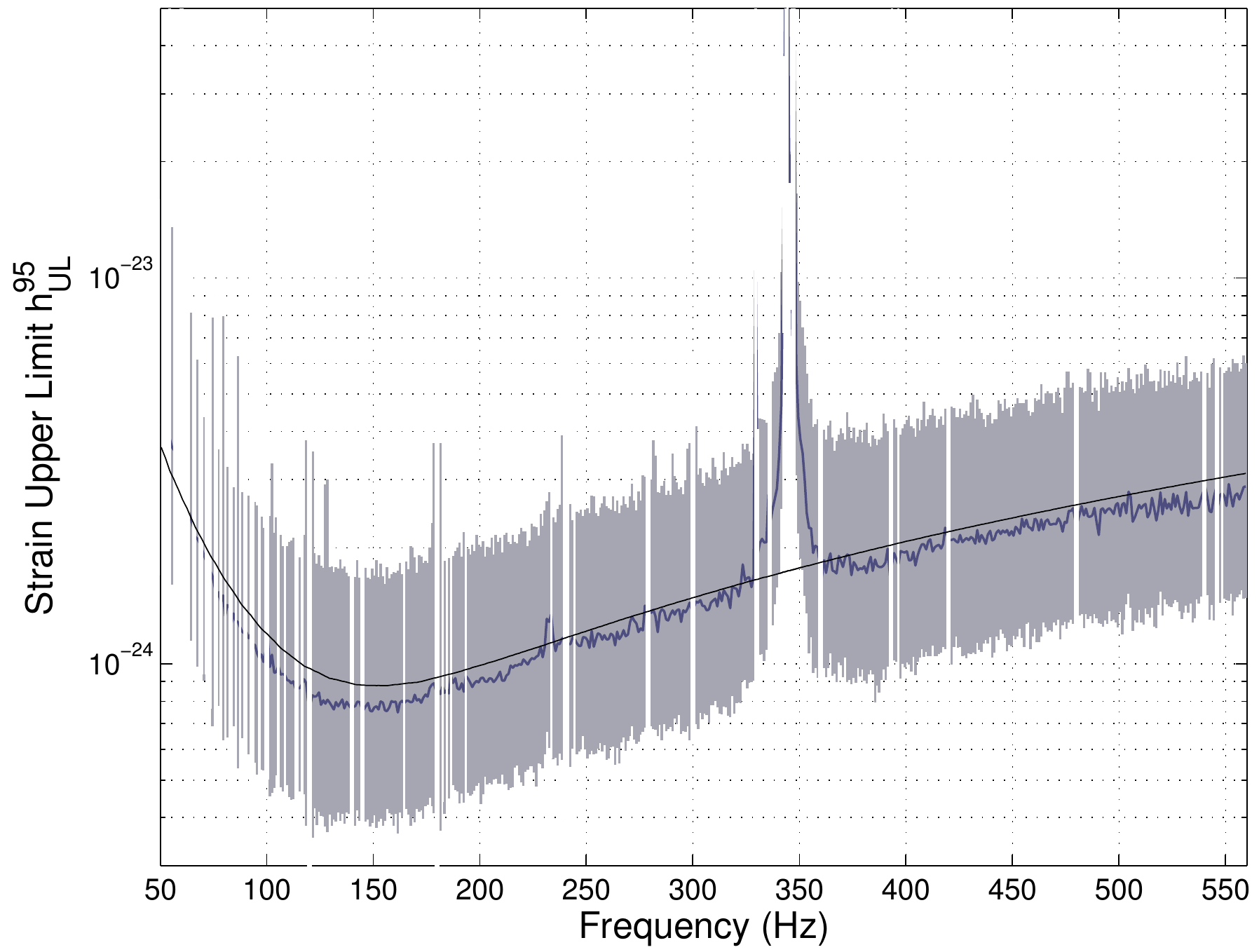}}
\caption{\label{fig:ULs} Gravitational wave strain $95\%$ upper limits
  for H1L1 data from 21--31 Aug 2007 for \subref{subfig:ULs} the
  standard search with flat priors on $\cos\iota$ and $\psi$ (left
  panel) and \subref{subfig:ULs_dir} the angle-restricted search with
  $\iota=44^\circ\pm6^\circ$ and $\psi = 234^\circ \pm 4^\circ$ (right
  panel). The grey region extends from the minimum to the maximum
  upper limit in each 1-Hz sub-band. The median upper limit
  in each sub-band is indicated by a solid, thick, blue-grey curve. The
  expected upper limit for Gaussian noise at the S5 design sensitivity
  is shown for comparison (solid, thin, black curve). Whited regions of the grey band indicate bands that have been excluded (due to known contamination or vetoed out bands). No upper limits are quoted in these bands.}
\end{figure*}

Bayesian upper limits are set using Eq.~\ref{eq:h0_UL} and an upper limit on $\ho$ is calculated for
every \cstext, yielding $2 \times 10^6$ results in each 1-Hz
sub-band. Figure~\ref{fig:ULs} shows the upper limits for our S5
dataset (21--31 Aug 2007) combining data from the LIGO H1 and L1
detectors. The grey band in Fig.~\ref{fig:ULs} stretches vertically
from the minimum to the maximum upper limit in each sub-band. The
solid grey curve indicates the expected value of the median $95\%$
upper limit for each sub-band given the estimated noise spectral
density in the selected data. The solid black curve indicates the
$95\%$ strain upper limit expected from Gaussian noise at the S5
design strain sensitivity and matches the median upper limit to within
$10\%$ in well-behaved (Gaussian-like) regions. The excursions from
the theoretical median, e.g. at $f\approx350$ Hz, are noise lines, as
discussed in Secs.~\ref{subsec:candidates} and \ref{subsec:veto}.

Figure~\ref{subfig:ULs} shows upper limits for the standard sideband
search, which adopts the electromagnetically measured values of
$\Porb$ and $\asini$ and flat priors on $\cos\iota$ and $\psi$
spanning their full physical range. Figure~\ref{subfig:ULs_dir} shows
upper limits for the sideband search using Gaussian priors on these
angles with preferred values of $\iota=44^\circ\pm6^\circ$ and
$\psi=234^\circ\pm3^\circ$ inferred from electromagnetic observations
of the \ac{ScoX1} jet. Section \ref{subsec:nuisance} describes the two
cases in more detail.

The minimum upper limit (i.e. minimised over each Hz band and shown as
the lower edge of the grey region in Fig.~\ref{fig:ULs}) between 120
and 150 Hz, where the detector is most sensitive, equals
$\hULn=6\times 10^{-25}$ with $95\%$ confidence for the standard
search, and $4\times 10^{-25}$ for the angle-restricted search. The variation agrees to within $5\%$ for both configurations of the search, for which the minimum and maximum vary from ${\sim} 0.5$ to ${\sim} 2$ times the median, respectively.

The strain upper limit $\hULn$ for the angle-restricted search in
Fig.~\ref{subfig:ULs_dir} is ${\sim}60\%$ lower than that of the
standard search in Fig. \ref{subfig:ULs} and the variation in span
between minimum and maximum within each sub-band is ${\sim}70\%$
narrower. Accurate prior knowledge of $\iota$ and $\psi$ reduces the
parameter space considerably. By constructing priors from the estimated values, the upper limits improve by a factor of 
1.5, though, this improvement can be applied independently of the search algorithm. 

%

\section{Discussion} \label{sec:discussion}

Accretion torque balance~\cite{Bildsten_1998} implies an upper limit
$\ho^\text{EQ}\leqslant7\times 10^{-26}$ at 150 Hz for \ac{ScoX1} (see
Eq.~\ref{eq:ScoX1_h0}). This sets the maximum expected strain at ${\sim}
6$ times lower than our angle-restricted upper limit ($4\times 10^{-25}$), assuming spin equilibrium as implied by torque balance. This is a conservative
limit. Taking the accretion-torque lever arm as the Alfv\'en radius
instead of the neutron star radius increases $\ho^\text{EQ}$ by a
factor of a few, as does relaxing the equilibrium assumption. Torque
balance may or may not apply if radiative processes modify the inner
edge of the accretion disk~\cite{PatrunoEA_2012}.

The sideband search upper limit can be used to place an upper limit on
the neutron star ellipticity $\epsilon$. We can express the
ellipticity $\epsilon$ in terms of $\ho$ by
\begin{eqnarray}
 \epsilon &=& \frac{5c^4}{8\pi G}\frac{D}{MR^2}\frac{\ho}{{\fspin}^2},
\end{eqnarray}
where $M$, $R$, and $\fspin$ are the mass, radius and spin frequency
of the star, respectively, and $D$ is its distance from
Earth~\cite{JKS1998}. Using fiducial values $\Mstar=1.4\Msol$ and $R=10$ km
and assuming $f=2\fspin$ (i.e. the principal axis of inertia is
perpendicular to the rotation axis), the upper limit $\epUL$ for
\ac{ScoX1} can be expressed as
\begin{equation}
 \epUL \leqslant 5\times 10^{-4} {\left(\frac{\hUL}{4\times 10^{-25}}\right)}{\left(\frac{f}{150 \text{ Hz}}\right)}^{-2} \left(\frac{D}{2.8 \text{ kpc}}\right).
\end{equation}
This is well above the ellipticities predicted by
most theoretical quadrupole generating mechanisms. Thermocompositional
mountains have $\epsilon \approx 9\times10^{-6}$ for $\lesssim 5\%$
lateral temperature variations in a single electron capture layer in
the deep inner crust or $0.5 \%$ lateral variations in charge-to-mass
ratio~\cite{UCB_2000}. Magnetic mountain ellipticities vary with the
\ac{eos}. For a pre-accretion magnetic field of $10^{12.5}$ G, one
finds $\epsilon \approx 2 \times 10^{-5}$ and $\epsilon \approx
6\times 10^{-8}$ for isothermal and relativistic-degenerate-electron
matter respectively~\cite{Vigelius_Melatos_2009, PriymakEA_2011}.
Equivalent ellipticities of $\epsilon {\sim} 10^{-6}$ are achievable by r-mode
amplitudes of a few times $10^{-4}$
\cite{OwenEA_1998,BondarescuEA_2007,LIGO_CasA_2010}. Our upper limit
approaches the ellipticity predicted for certain exotic equations of
state \cite{JMD_Owen_2013, GlampedakisEA_2012, Owen_2005}.

The sideband search presented here is restricted to $\Tobs = 10$ days
due to current limitations in the understanding of spin wandering, i.e. the fluctuations in the neutron star spin frequency due to a time varying accretion torque. The 10-day restriction follows from the accretion torque fluctuations inferred
from the observed X-ray flux variability, as discussed in Section
IV. B. in~\cite{SammutEA_2014}. Improvements in understanding of this
feature of phase evolution and how to effectively account for it could
allow us to lengthen $\Tobs$ and hence increase the sensitivity of the
search according to $\ho\propto\Tobs^{-1/2}$. Pending that, results from multiple 10-day stretches could be incoherently combined, however, the unknown change in frequency between observations must be accounted for. Sensitivity would also increase if data from additional comparably sensitive detectors are included, since $\ho \propto N_{\text{det}}^{-1/2}$, where $N_{\text{det}}$ is
the number of detectors, without significantly increasing the
computational cost~\cite{Prix_2007}.

\section{Conclusion} \label{sec:conclusion}

We present results of the sideband search for the candidate
gravitational wave source in the \ac{LMXB} \ac{ScoX1}. No evidence was
found to support detection of a signal with the expected waveform. We
report $95\%$ upper limits on the gravitational wave strain $\hULn$
for frequencies $50 \leqslant f \leqslant 550$ Hz. The tightest upper
limit, obtained when the inclination $\iota$ and gravitational wave
polarisation $\psi$ are known from electromagnetic measurements, is
given by $\hULn\approx 4\times 10^{-25}$. It is achieved for $120 \leqslant f \leqslant 150$
Hz, where the detector is most sensitive. The minimum upper limit for
the standard search, which assumes no knowledge of source orientation (i.e. flat priors on $\iota$ and $\psi$), is
$\hULn=6\times 10^{-25}$ in this frequency range. The median upper
limit in each 1-Hz sub-band provides a robust and representative
estimate of the sensitivity of the search. The median upper limit at 150 Hz was $1.3\times 10^{-24}$ and $8\times 10^{-25}$ for the standard and angle-restricted searches respectively.

The results improve on upper limits set by previous searches directed
at \ac{ScoX1} and motivates future development of the search. The improvement in results is achieved using only a 10-day coherent observation time, and with modest computational expense. Previously, using roughly one year of coincident S5 data, the radiometer search returned a median $90\%$ root-mean-square strain upper limit of $h_\text{rms}^{90}\approx7\times10^{-25}$ at 150 Hz \cite{LIGO_S5_ScoX1stoch_2011}, which converts to $h_0^{95} \approx 2\times10^{-24}$ \cite{strain, rad, Messenger2010}.

The first all-sky search for continuous gravitational wave sources in
binary systems using the TwoSpect algorithm has recently reported results using ${\sim} 1.25$ years of S6
data from the LIGO and VIRGO detectors ~\cite{Goetz_Riles_2011, LIGO_TwoSpect_2014}. Results of an adapted version of the analysis directed at \ac{ScoX1} assuming the electromagnetically measured values of $\Porb$ and $\asini$ was also reported together with the results of the all-sky search. Results of this analysis are comparable in sensitivity to the sideband
search. Results for the \ac{ScoX1} directed search were restricted to the frequency band $20
\leqslant f \leqslant 57.25$ Hz due to limitations resulting from
1800-s \acp{SFT}.

We have shown that this low computational cost, proof-of-principle analysis, applied to only 10 days of data, has provided the most sensitive search for gravitational waves from Sco X-1. The computational efficiency and relative sensitivity of this analysis over relatively short coherent time-spans makes it an appealing search to run as a first pass in the coming second-generation gravitational wave detector era. Running in low-latency with the capability of providing updated results on a daily basis for multiple LMXB systems would give the first results from continuous wave searches for sources in known binary systems. With a factor 10 improvement expected from advanced detectors, and the sensitivity of semi-coherent searches improving with the fourth root of the number of segments and, for our search, also with the square root of the number of detectors, we can hope for up to a factor ${\sim}$30 improvement for a year long analysis with 3 advanced detectors. This would place the sideband search sensitivity within reach of the torque-balance limit estimate of the \ac{ScoX1} strain (Eq. ~\ref{eq:ScoX1_h0}) in the most sensitive frequency range, around 150 Hz. However, the effects of spin wandering will undoubtedly weaken our search and impose a larger trials factor to our detection statistic, therefore increasing our detection threshold. Efficient analysis methods that address spin wandering issues to allow longer coherent observations or combine results from separate observations should improve the sensitivity of the search, enhancing its capability in this exciting era.

\begin{acknowledgments}
The authors gratefully acknowledge the support of 
the United States National Science Foundation (NSF) for the construction and operation of the LIGO Laboratory,
the Science and Technology Facilities Council (STFC) of the United Kingdom, 
the Max-Planck-Society (MPS), and the State of Niedersachsen/Germany 
for support of the construction and operation of the GEO600 detector,
the Italian Istituto Nazionale di Fisica Nucleare (INFN) and 
the French Centre National de la Recherche Scientifique (CNRS)
for the construction and operation of the Virgo detector
and the creation and support of the EGO consortium. 
The authors also gratefully acknowledge research support from these agencies as well as by 
the Australian Research Council,
the International Science Linkages program of the Commonwealth of Australia,
the Council of Scientific and Industrial Research of India, 
Department of Science and Technology, India,
Science \& Engineering Research Board (SERB), India,
Ministry of Human Resource Development, India,
the Spanish Ministerio de Econom\'ia y Competitividad,
the Conselleria d'Economia i Competitivitat and Conselleria d'Educaci\'o, Cultura i Universitats of the Govern de les Illes Balears,
the Foundation for Fundamental Research on Matter supported by the Netherlands Organisation for Scientific Research, 
the Polish Ministry of Science and Higher Education, 
the FOCUS Programme of Foundation for Polish Science,
the European Union,
the Royal Society, 
the Scottish Funding Council, 
the Scottish Universities Physics Alliance, 
the National Aeronautics and Space Administration, 
the Hungarian Scientific Research Fund (OTKA),
the Lyon Institute of Origins (LIO),
the National Research Foundation of Korea,
Industry Canada and the Province of Ontario through the Ministry of Economic Development and Innovation, 
the National Science and Engineering Research Council Canada,
the Brazilian Ministry of Science, Technology, and Innovation,
the Carnegie Trust, 
the Leverhulme Trust, 
the David and Lucile Packard Foundation, 
the Research Corporation, 
and the Alfred P. Sloan Foundation.
The authors gratefully acknowledge the support of the NSF, STFC, MPS, INFN, CNRS and the
State of Niedersachsen/Germany for provision of computational resources. 

This article has LIGO Document No. LIGO-P1400094.
\end{acknowledgments}

\bibliography{references}

\begin{thebibliography}{58}
\bibitem [{str()}]{strain}%
  \BibitemOpen
  \href@noop {} {}\bibinfo {note} {The wave strain $\ho$ commonly used in
  continuous wave searches can be related to $h_{\text{rms}}$ by $\ho \approx
  2.43 h_\text{rms}$, and for this search $h_0^{95} \approx 1.2
  h_0^{90}$.}\BibitemShut {Stop}%
\end{thebibliography}

\appendix
\section{Manual veto} \label{ap:veto}

The eight candidates surviving the automated $4M$ veto, listed in Table~\ref{tab:noise_veto}, were
followed up manually. The manual follow-up of these candidates is presented here in more detail. Both the automated and manual veto stages were tested on software injected signals and simulated Gaussian noise to ensure signals were not accidentally vetoed. The tests showed that the vetoes are conservative.

\begin{figure*}
\subfigure{\includegraphics[width=.42\textwidth]{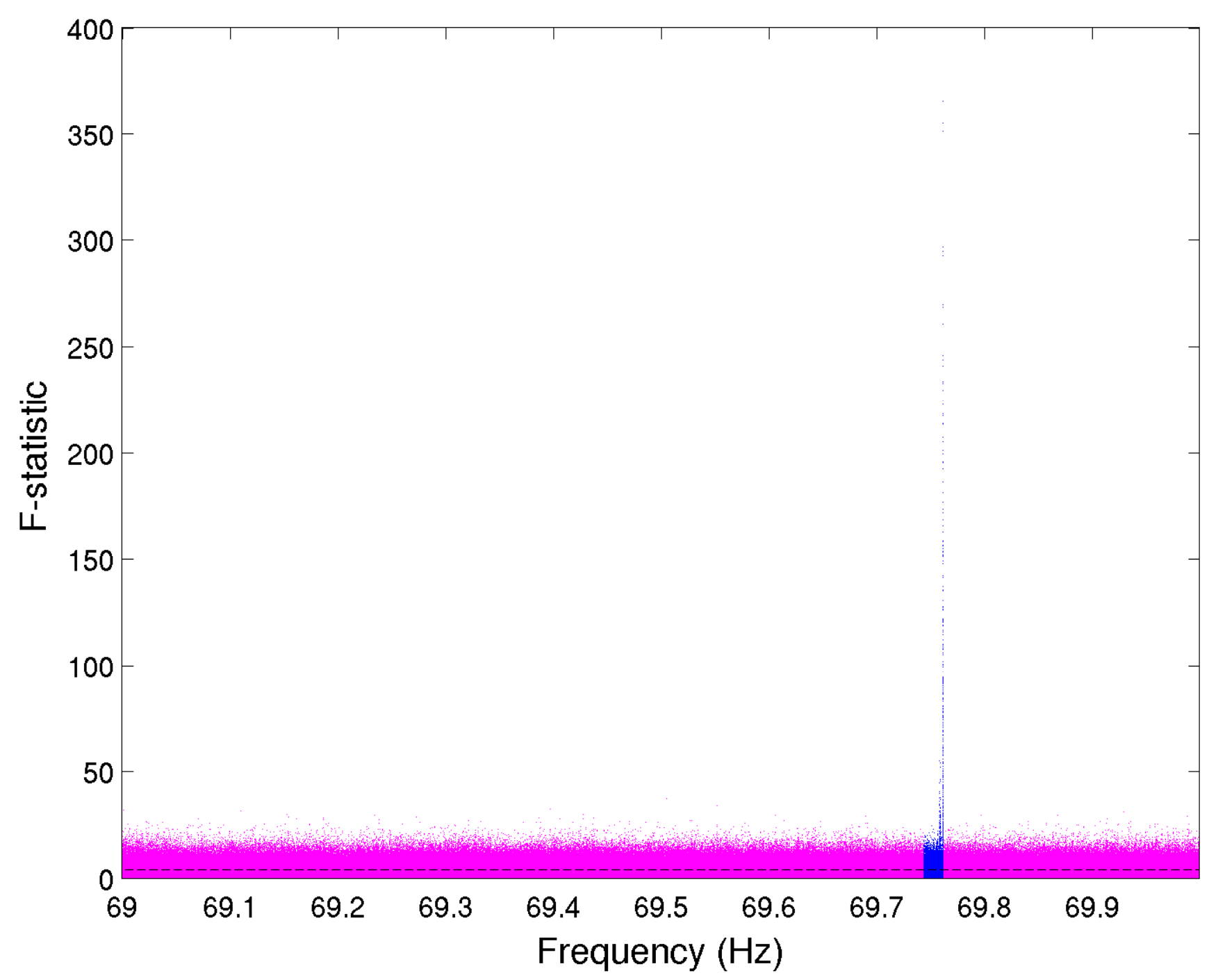}}
\subfigure{\includegraphics[width=.42\textwidth]{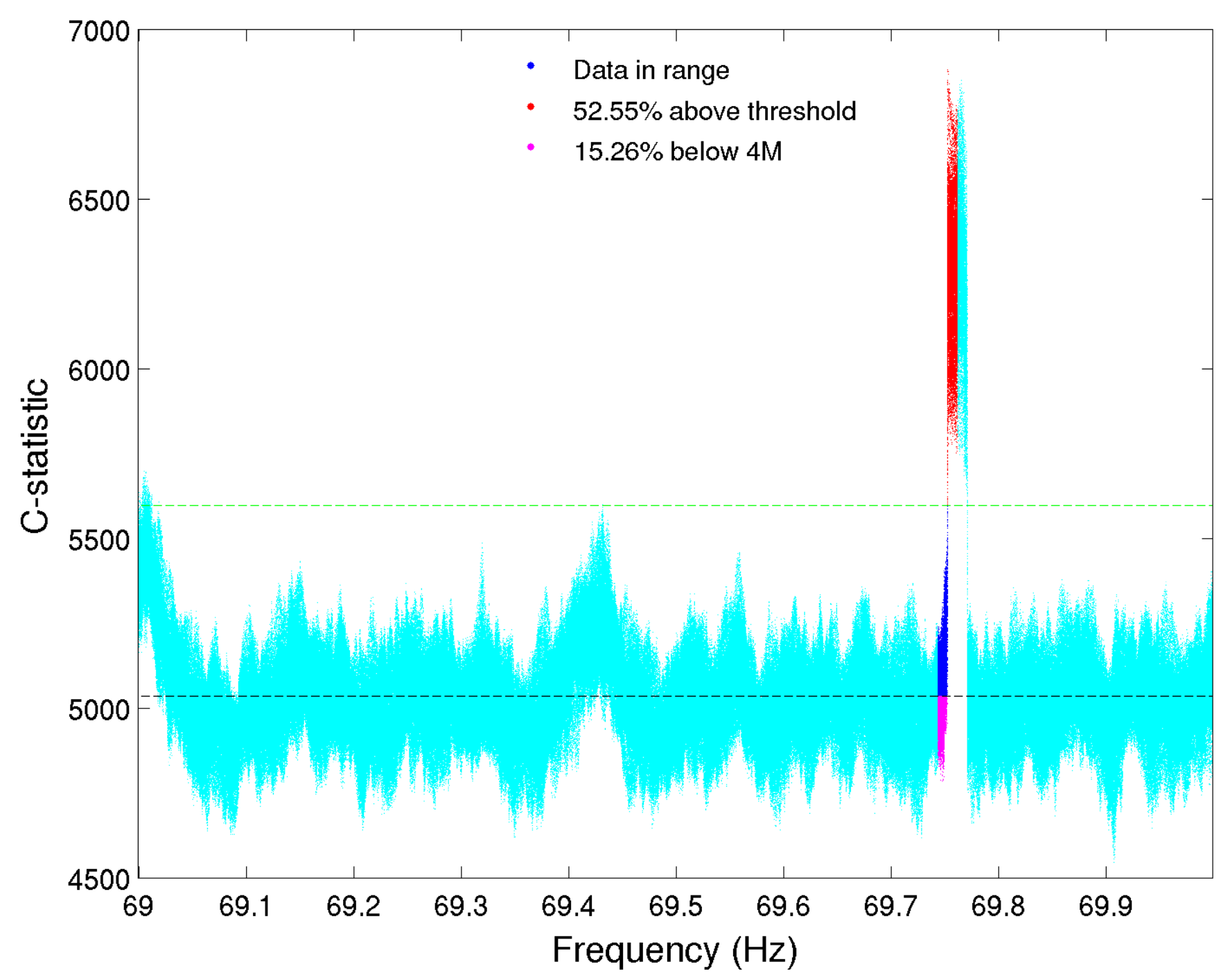}}
\caption{\label{fig:data_veto_line}(colour online) \fstext (left, magenta) and \cstext (right, cyan) versus frequency for Hz sub-band beginning at 69 Hz containing a candidate surviving the $4M$ veto which is attributed to a noise line. The frequency range used to determine the veto is highlighted (blue points). Points in the \cstext veto region are further highlighted (in red) if they exceed the threshold $\cstar_\kappa$ and (in pink) if they fall below the expectation value of the noise $4M$. The horizontal black dashed line indicates the expected value for noise ($\fstat=4$, $\cstat=4M$). The threshold value $\cstar_\kappa$ is also indicated on the \cstext plots by a horizontal green dashed line. The percentage of \cstexts falling above $\cstar_\kappa$ or below $4M$ is quoted in the legend in each \cstext panel.}
\end{figure*}

\begin{figure*}
\subfigure{\includegraphics[width=.42\textwidth]{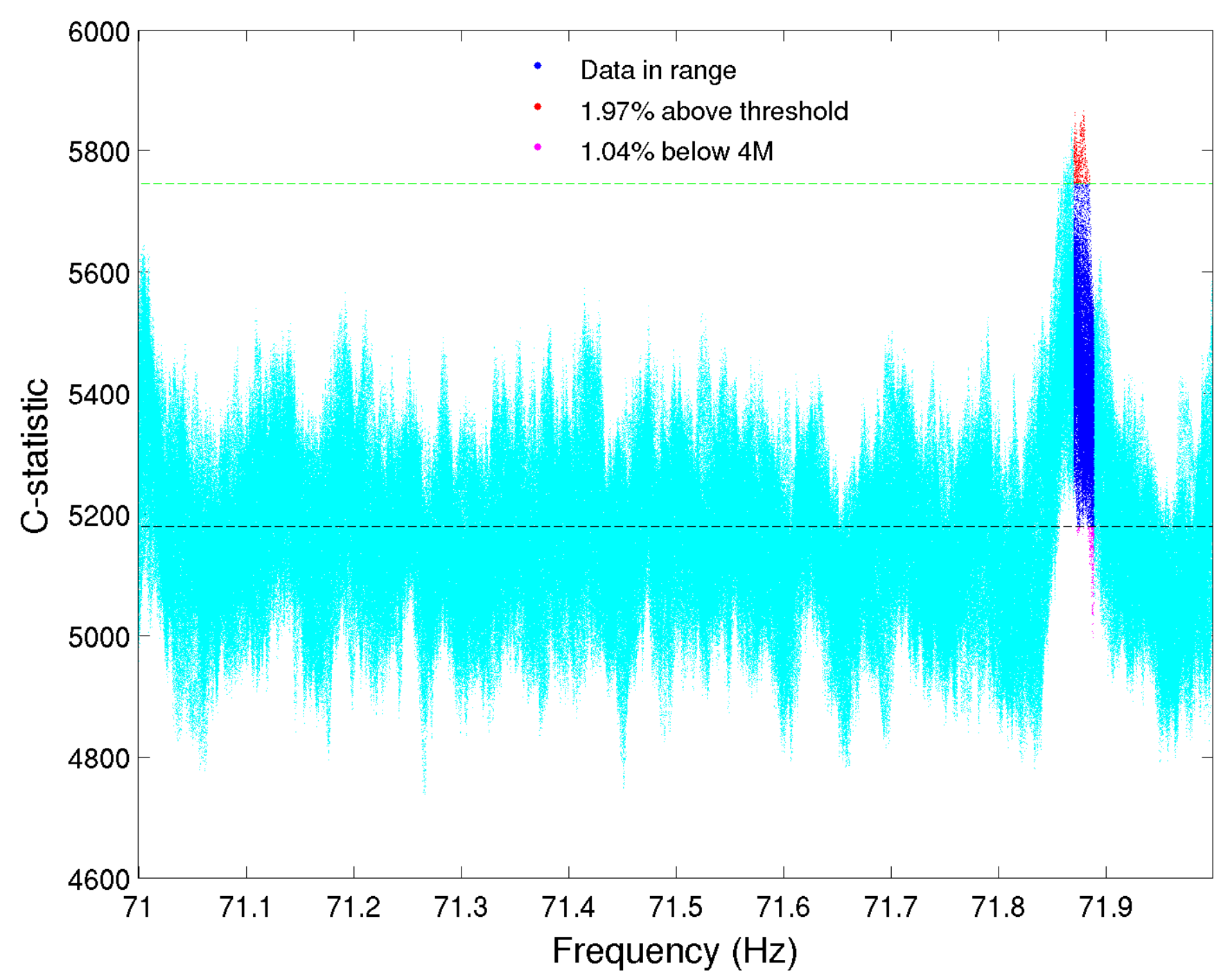}}
\subfigure{\includegraphics[width=.42\textwidth]{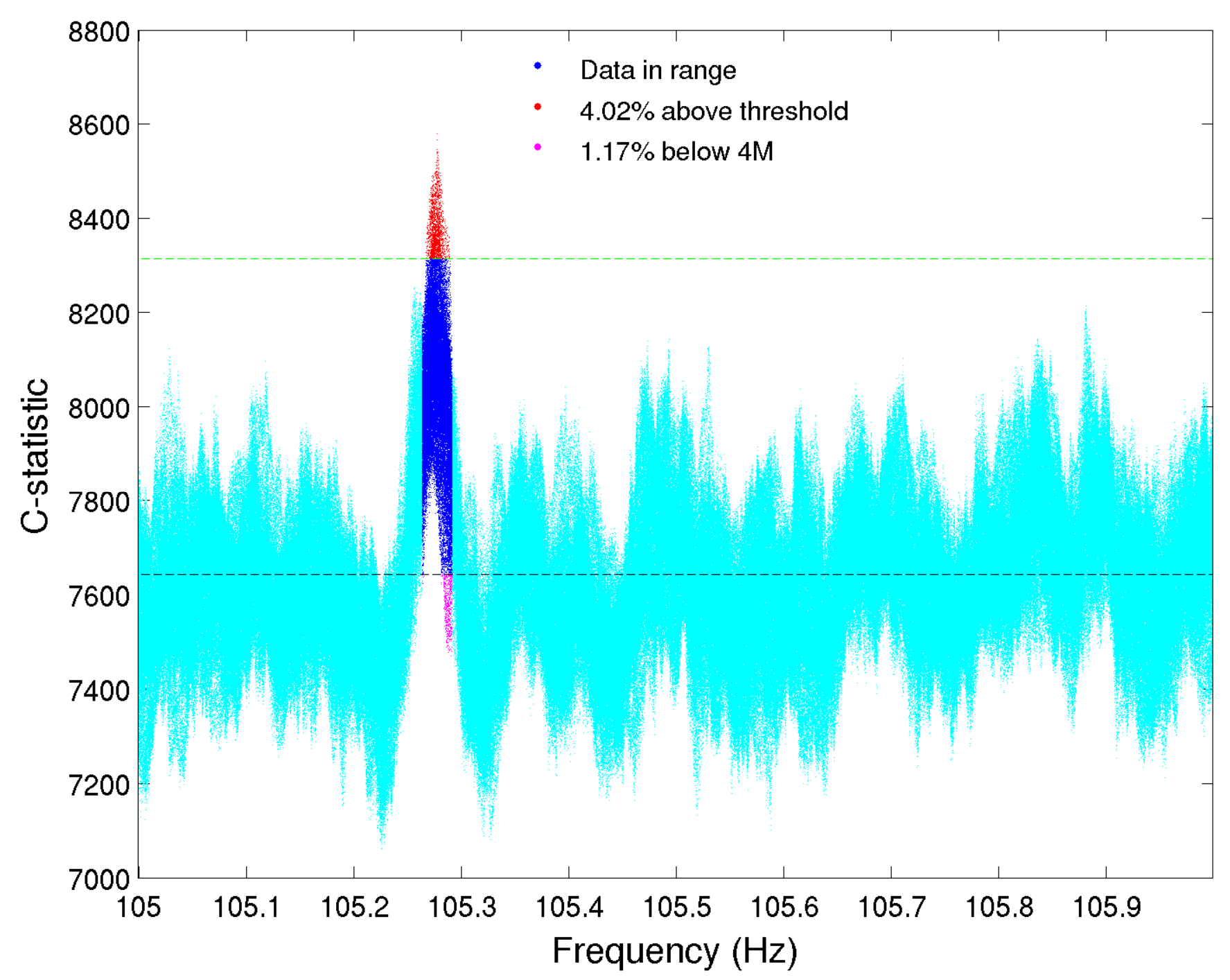}}
\subfigure{\includegraphics[width=.42\textwidth]{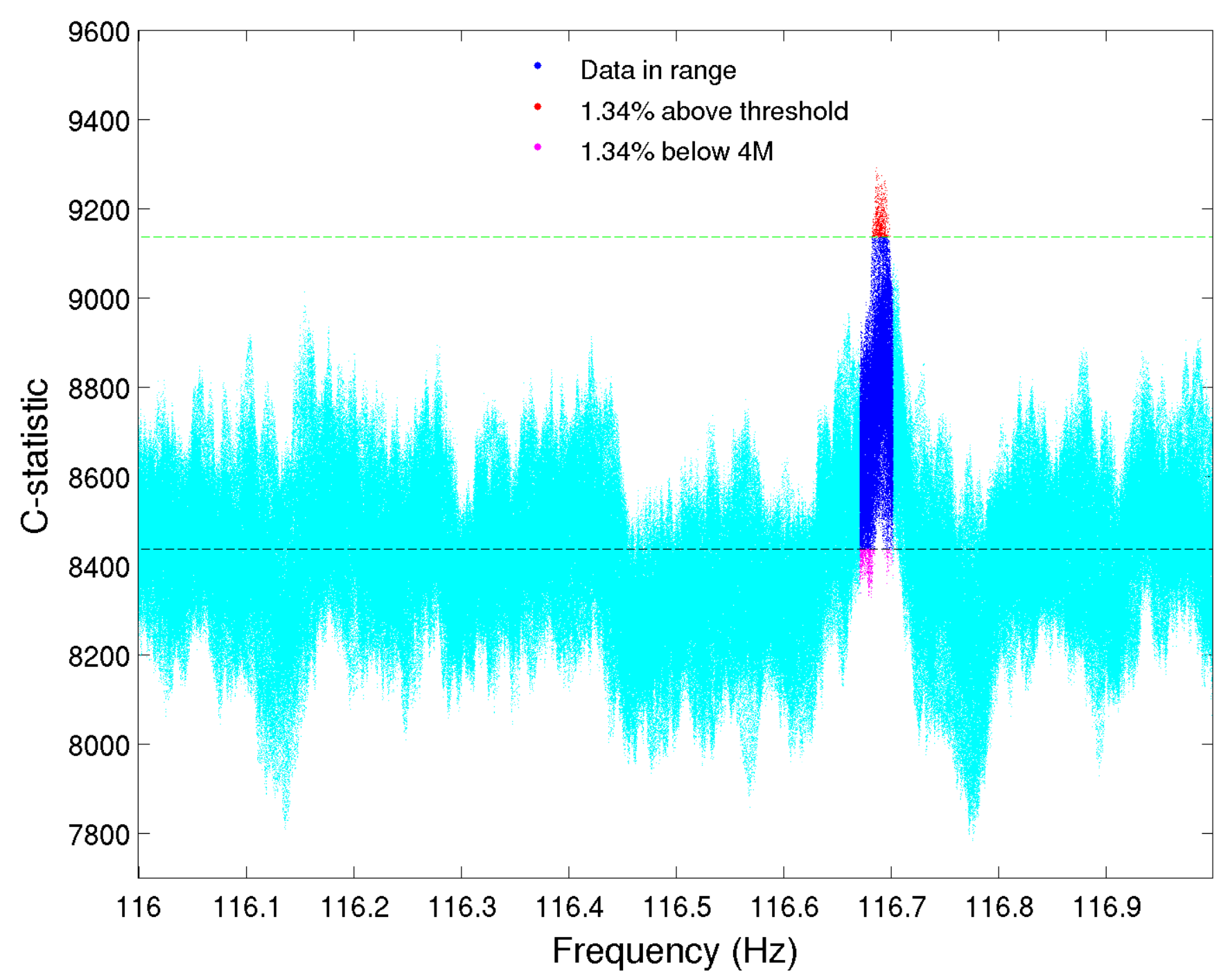}}
\subfigure{\includegraphics[width=.42\textwidth]{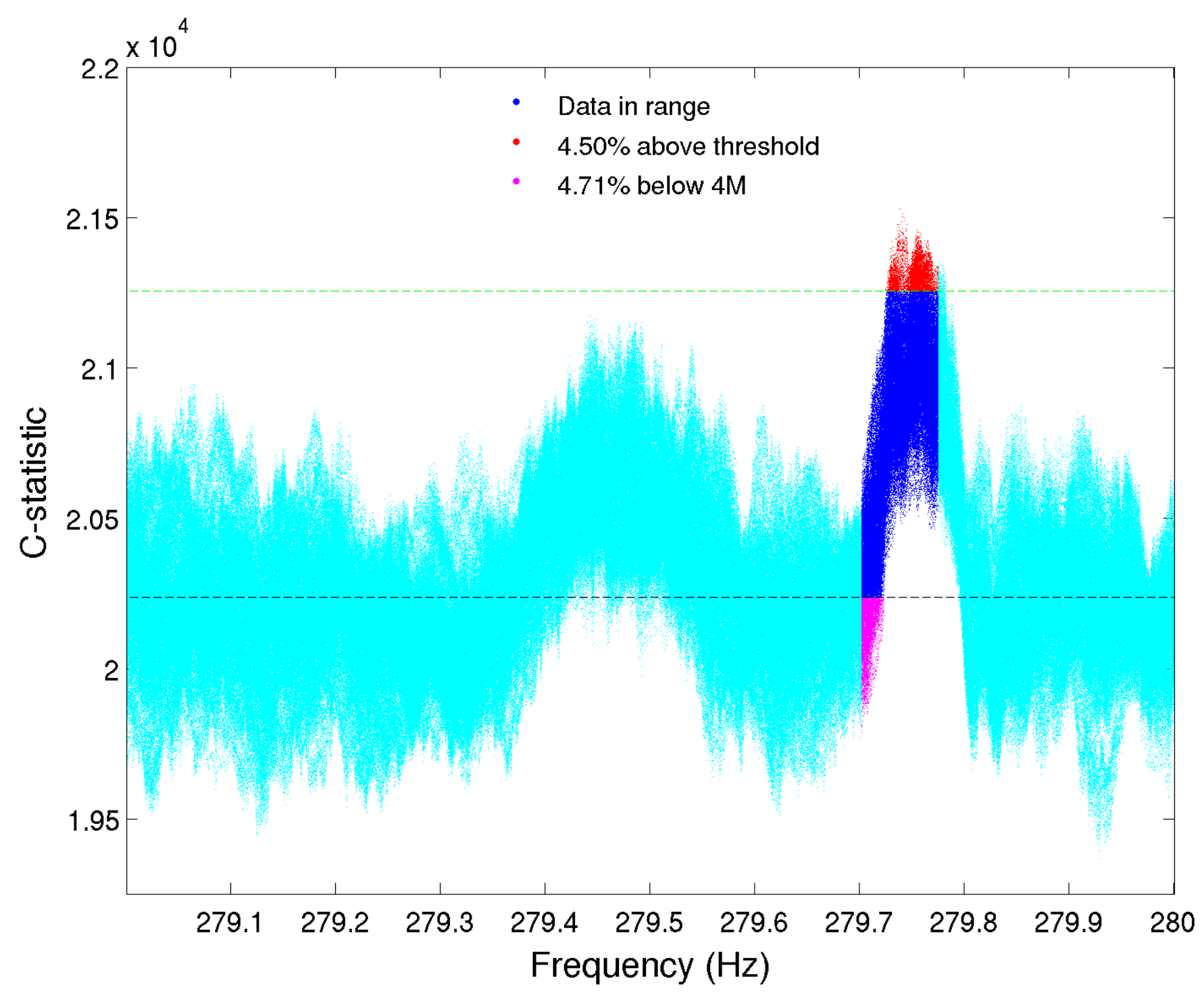}}
\caption{\label{fig:data_veto_noise}(colour online) As for Figure \ref{fig:data_veto_line} but for sub-bands beginning at 71, 105, 116 and 279 Hz containing candidates surviving the $4M$ veto with features not consistent with a signal, which are attributed to noise.}
\end{figure*}

\begin{figure*}
\subfigure{\includegraphics[width=.5\textwidth]{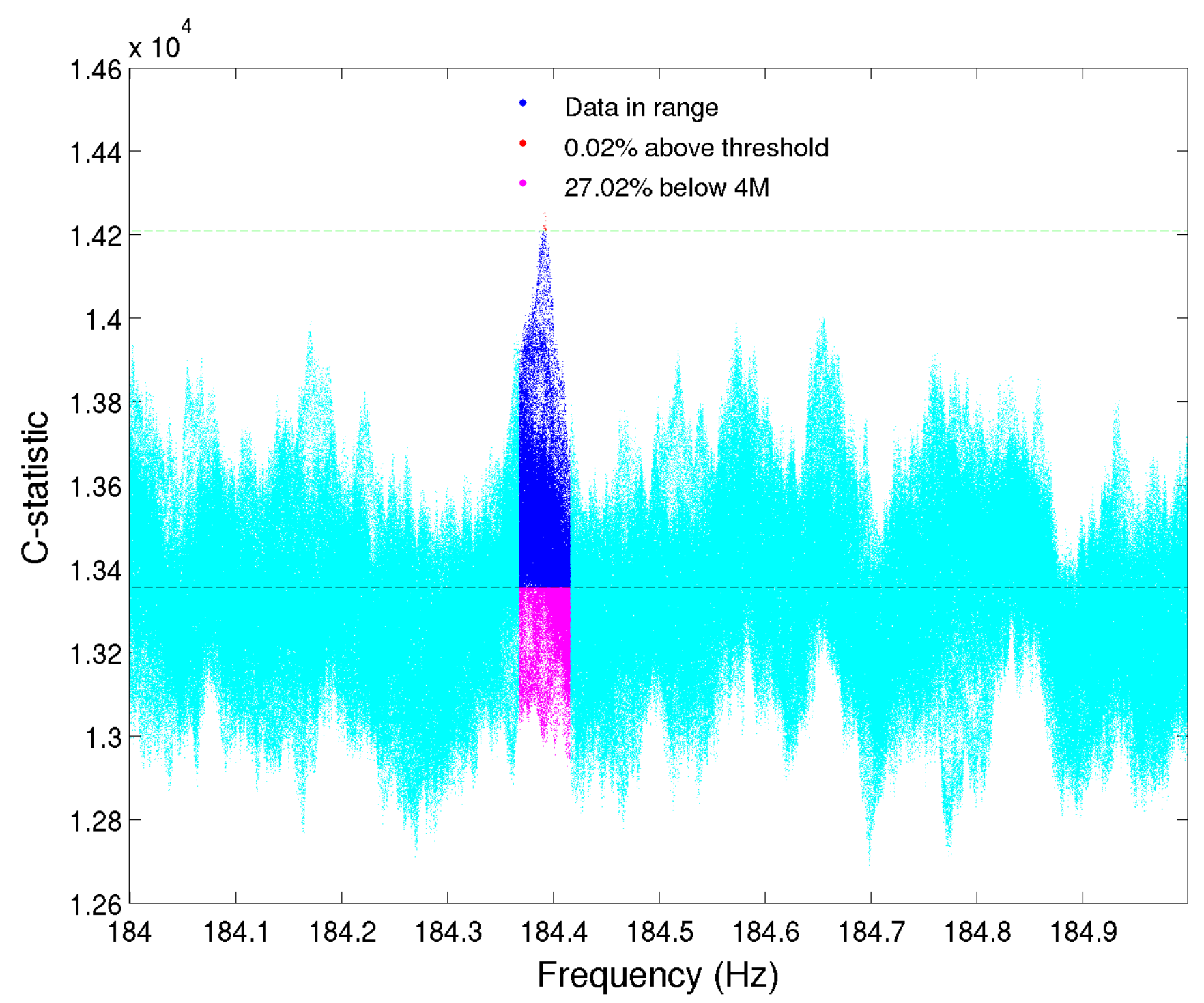}}
\subfigure{\includegraphics[width=.5\textwidth]{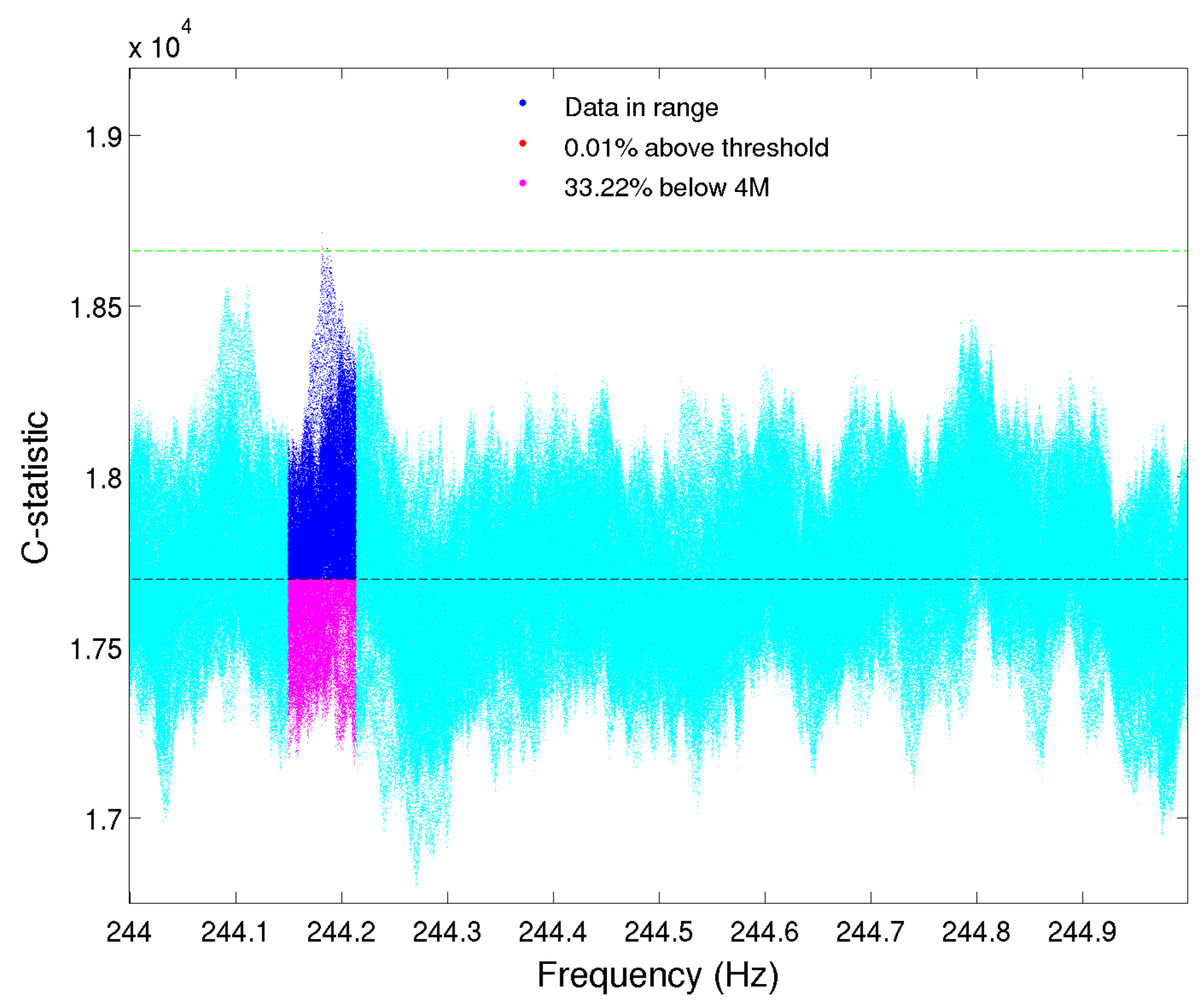}}
\subfigure{\includegraphics[width=.5\textwidth]{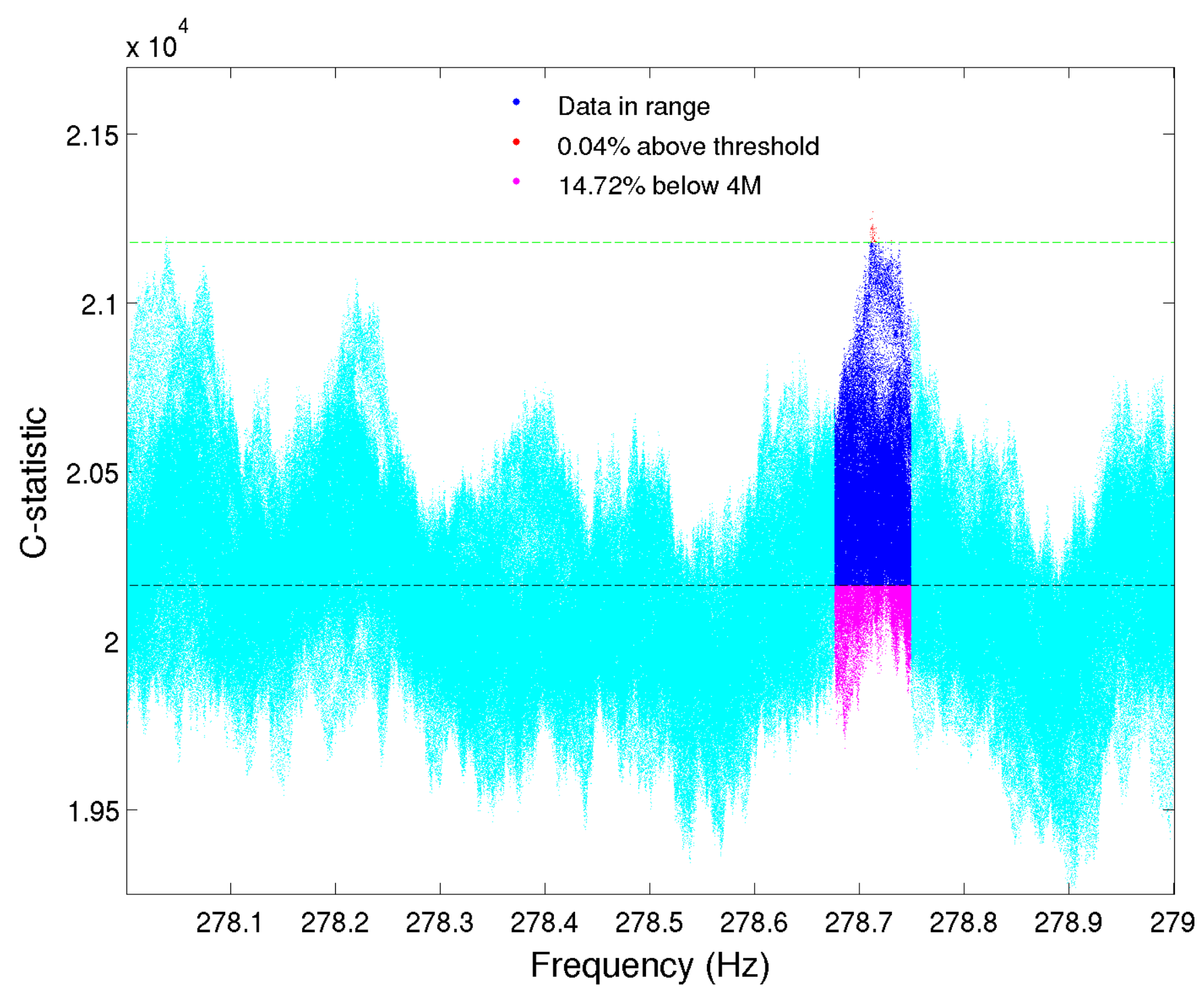}}
\caption{\label{fig:data_veto_unclear}(colour online) As for Figure \ref{fig:data_veto_line} but for sub-bands beginning at 184, 244 and 278 Hz that survive the $4M$ veto which are consistent with false alarms expected from noise.}
\end{figure*}

Figures \ref{fig:data_veto_line}, \ref{fig:data_veto_noise} and \ref{fig:data_veto_unclear} display the output (\fstext in magenta and \cstext in cyan) for the 1-Hz sub-bands containing the candidates surviving the $4M$ veto. The frequency range used
for the veto is highlighted in blue in each plot. Some \cstexts in
this region are further highlighted in red if they exceed the
threshold $\cstar_\kappa$ or magenta if they fall below $4M$. The
expected values ($\fstat=4$ and $\cstat=4M$) are indicated by solid
black-dashed horizontal lines. The threshold $\cstar_\kappa$ is
indicated by a green horizontal dashed line in each of the \cstext
plots.

Figure~\ref{fig:data_veto_line} displays the output of the sub-band starting at 69 Hz, containing a candidate judged to arise from a noise line. The line is clearly evident in the \fstext (left hand panel). The sideband signal
targeted by this search will be split over many \fstext bins due to the modulation caused by the motion of the source in its binary orbit. The signal is not expected to be contained in a single bin. The veto should
automatically rule out single-bin candidates such as this one, however the veto fails
to reject this candidate because $f_{\text{max}}$ (where the veto band
is centered) falls closer to one end of the contaminated region rather
than the centre. In this special scenario the veto picks up several
bins with $\cstat<4M$ from just outside the contaminated region (where
the noise is ``normal'') so the candidate survives. Visual inspection
is important in these cases and shows clearly that the candidate could
not result from a signal.

Figure \ref{fig:data_veto_noise} shows the \cstext output for the other candidate sub-bands attributed to noise. The features visible in the \cstext can
be ruled out as originating from an astrophysical signal since the
fraction of bins above $4M$ is too large compared to what would be
expected from such a signal with the same apparent \ac{SNR}. We would
expect the frequency bins in between sidebands to drop down to values
of $\cstat {\sim} 4M$, resulting in a consistent noise floor even around the candidate ``peak''. The elevated noise floor around the peaks is not consistent with an expected signal. Similar features can be seen in each of the sub-bands starting at 71, 105, 116 and 279 Hz.

Figure~\ref{fig:data_veto_unclear} presents the candidates in the 184, 244 and 278
Hz sub-bands, which are consistent with false alarms expected from noise. The candidate in the 184 Hz sub-band has a
healthy fraction ($26\%$) of bins with $\cstat<4M$ and resembles the
filled dome with consistent noise floor expected from a signal (unlike the examples in
Fig. \ref{fig:data_veto_noise}), although it is slightly pointier. At
$f=244$ Hz, $33\%$ of bins have $\cstat<4M$ but the \cstext pattern is
multimodal and less characteristic of a signal. The candidate peak is
comparable in amplitude to several other fluctuations within the
sub-band, possibly indicating a contaminated (non-Gaussian)
noise-floor. Similar remarks apply to $f=278$ Hz, especially
consideration of the noise-floor fluctuations. Additionally, the
candidate at 278 Hz also coincides with a strong, single-bin spike in
the \fstext at 278.7 Hz.


\end{document}